# Electron Dynamics in Films Made of Transition Metal Nanograins Embedded in $SiO_2$:

# Infrared Reflectivity and Nanoplasma Infrared Resonance

Néstor E. Massa*
Laboratorio Nacional de Investigación y Servicios en Espectroscopía Optica-CEQUINOR, Universidad Nacional de La Plata, C.C. 962 , 1900 La Plata, Argentina,

Juliano C. Denardin,
Departamento de Física, Universidad de Santiago de Chile, Av. Ecuador 3493, Santiago, Chile,

Leandro M. Socolovsky,
Instituto de Tecnologías y Ciencias de la Ingeniería,
Universidad de Buenos Aires, Av. Paseo Colón 850, Buenos Aires, Argentina,

Marcelo Knobel,
Instituto de Física, "Gleb Wataghin", Universidade Estadual de Campinas, 13083-970, Campinas, SP, Brazil,

XiXiang Zhang,
Physics Department and Institute of Nanoscience and Technology, Hong Kong, University of Science and Technology, Clear Water Bay, Kowloon, Hong Kong, China.

- E-mail address: : neemmassa@gmail.com

# Abstract

We report on near normal infrared reflectivity spectra of ~550 nm thick films made of cosputtered transition metal nanograins and $SiO_2$ in a wide range of metal fractions. $Co_{0.85}(SiO_2)_{0.15}$,with conductivity well above the percolation threshold has a frequency and temperature behavior according to what it is find in conducting metal oxides. The electron scattering rate displays an unique relaxation time characteristic of single type of carriers experiencing strong electron-phonon interactions. Using small polaron fits we individualize those phonons as glass vibrational modes. $Ni_{0.61}(SiO_2)_{0.39}$, with a metal fraction closer to the percolation threshold, undergoes a metal-non metal transition at ~77 K. Here, as it is suggested by the scattering rate nearly quadratic dependence, we identify two relaxation times (two carrier contributions) associated to a Drude mode and a mid-infrared overdamped band, respectively. Disorder induced, the mid-infrared contribution drives the phase transition by thermal electron localization. $Co_{0.51}(SiO_2)_{0.49}$ has the reflectivity of an insulator with a distinctive band at ~1450cm$^{-1}$ originating in electron promotion, localization, and defect induced polaron formation. Angle dependent oblique reflectivity of globally insulating $Co_{0.38}(SiO_2)_{0.62}$, $Fe_{0.34}(SiO_2)_{0.66,}$ and $Ni_{0.28}(SiO_2)_{0.72}$, reveals a remarkable resonance at that band threshold. We understand this as due to the excitation by normal to the film electric fields of defect localized electrons in the metallic nanoparticles. At higher oblique angles, this localized nanoplasma couples to $SiO_2$ longitudinal optical Berreman phonons resulting a band peak softening reminiscent to the phonon behavior undergoing strong electron-phonon interactions. Singular to a globally insulating phase, we believe that this resonance might be a useful tool for tracking metal-insulator phase transitions in inhomogeneous materials.

## Introduction

In the recent past there have been many attempts to describe the conduction mechanisms in films made of metallic grains in an insulating matrix films. Besides being mainly theoretical,[1] most experiments were carried out analyzing transport and magnetic measurements.[2] They have been mostly addressed to mechanisms behind transport and carrier scattering involving the Giant Hall Effect [3,4] which depends on a combination of nanoparticle diameter, interparticle distance, and size distribution.[5] There have also been a plethora of theoretical work addressing the electronic transport in this kind of systems.[6]

Here we attempt to improve the understanding of spatially inhomogeneous systems with inherent complexity[7] by expanding our knowledge on the electron dynamics in well characterized films made of transition metal nanograins cosputtered with $SiO_2$. Phonons are expected to be electron dressed as in devices running from single molecular transistors to electron-phonon quantum dot interactions.[8]

We have carried out temperature dependent far infrared measurements with techniques that allow a quantitative and reliable evaluation of samples with different metal transition volume fractions. We attribute significance to our study because it also reveals a behavior close what it is generally found in optical measurements for bulk conducting transition metal oxides where quenched disorder provides ground for rather complex behaviors.[9]

Examples of earlier spectroscopic work may be traced to absorption measurements by Devaty and Sievers[10] on the study of ~100 Å Ag particles embedded in an insulating gelatin matrix. The optical properties of several volume fractions of Ag small particles in KCl were also studied by Cummings et al [11]; Kim and Tanner reported on the infrared response of aluminum small particles[12] and closer to our reflectivity work Noh et al[13] studied percolation through optical properties of Ni-MgO composites. Having as common ground sample characterization by transmission electron microscopy (TEM) those measurements only achieve qualitative agreement when analyzed with the Maxwell-Garnet dielectric theory[14] or effective medium approaches [15,16,17]. Room temperature transmission measurements of cosputtered Ag-$SiO_2$ and Au-$SiO_2$ granular films with a wide metal fraction were studied by Cohen et al[18]. It was then concluded that metal particles as small as 20 Å have optical constants that do not differ significantly from those of the bulk metals. More recently, absorption and photoluminescense properties of sol-gel granular transition metals in $SiO_2$ matrices have been reported by Yeshchenko et al.[19,20]

Spatial homogeneity is at the basis of the understanding on the intrinsic behavior of transition metal oxides, and in particular, the metal to insulator transitions found in highly correlated oxides. Ranging from high Tc superconductors to manganites they show a plethora of properties related to several competing states due to subtle phase separations. These materials reveal intrinsic inhomogeneities being substantiated in manganites as nanoscale granularity.[22] This might also be concomitant to cuprates where phase competition might be tied to oxygen vacancies distributed at random in the lattice inhomogeneous patterns reflecting disorder and distortion. Then, our results also aim to add a critical experimental view of the nature of metal–insulator transition in oxides as our results are from known samples made of components providing completely disordered environment. I.e., to contrast the way that it is currently understood the clean limit, against the relevance of incorporating inhomogeneities, i.e., quenched disorder, by which two states, insulating and metallic, coexist nearly degenerated.[9] It is to also note that glassy behavior is claimed near a metal-insulator transition.[11]

As a step toward that aim we show that our films, with plain glass network disorder as the main characteristic in the micro-scale, have a rather straight macro-scale response to tools regularly used in the field of infrared spectroscopy. In other words, what we analyze is the film response of nanoscale metallic objects in an macroscopic insulating matrix at the far-mid infrared (wavelength: ~ 2 to 1000 microns; wavenumbers ~ 11000 to 20 $cm^{-1}$). We obtain spectra that are analogous to those regularly found in oxides in which we verify, as for transport measurements, that in a suitable percolating network polarons are formed.

We selected for our study films with three different transition metal fractions whose response is assimilated to a metal oxide, to an intermediate case with lower conductivity, and to an insulator respectively. In the first case the temperature dependent specular reflectivity of $Co_{0.85}(SiO_2)_{0.15}$ is found to have similar features as encountered in the current literature for conducting oxides. They have a Drude component, phonons mostly carrier screened at far infrared frequencies, and a long tail that extents toward near infrared associated with hopping electron conductivity, and the presence of strong electron-phonon interactions. $Ni_{0.61}(SiO_2)_{0.39}$ represents the intermediate case that it is interesting because at about these metal transition concentrations the percolating network is reduced. Being near the percolation threshold, they have giant enhancements in the two contributions to the Hall resistivity.[3,4] The temperature dependent near normal reflectivity shows that $Ni_{0.61}(SiO_2)_{0.39}$ is a system in which the relative reduction in the number of carriers allows less screened phonon bands on the top of a continuum. In this example the spectra main features correspond to a weaker Drude Lorentzian, centered at zero frequency, and to a wide mid-infrared overdamped oscillator. This is understood as a profile of convoluted states traced to transitions of localized self-trapped charges that can be photoexcited to glass defects such as dangling bonds, these playing the role of unintended doping.

The third case is dedicated to samples in which most of the conducting critical paths are truncated and for which the nanoparticles are better defined. They now have infrared spectra with well defined vibrational bands and a sharp threshold at ~1450 $cm^{-1}$. Most remarkable, it is a distinctive resonant sharp peak found for the p-polarized angle dependent specular reflectivity. Found at the threshold edge of the band associated with electron promotion to higher energies, it is interpreted as due to a near surface defect localized electrons that shaking against the positive background create the electric dipole detected in the infrared. These last measurements unequivocally individualize that edge, also found in all semiconducting oxides, as having an electronic origin.

We reason that all these measurements may also be thought as a helpful first experimental approach describing films made of real metal nanoparticles in an insulating environment yielding grounds for plasmon circuit design.[23]

It is important to note that the overall properties discussed in the present manuscript apply to all our films regardless on the metal transition used in making the films. Our samples are superparamagnetic down to 4 K.

## Sample Preparation and Characterization

Our films made of $TM_x(SiO_2)_{1-x}$ ($0 < x < 1$; TM= Fe, Co, Ni) ~550 nm thick have been made by magnetron cosputtering. Films were deposited on glass and kapton substrates kept at room temperature. The chamber pressure before sputtering was kept at $10^{-7}$ tor. During the deposition, while the substrate was rotated to assure composition uniformity, the chamber was kept at 5 mTorr in an Ar atmosphere.

The metal volume fractions were controlled by changing relative sputtering rates and subsequently verified by energy dispersive X-ray spectroscopy (Philips EDAX XL30) which yielded ~7% uncertainty for the film metal fraction.

We have also characterized these samples by transmission electron microscopy (TEM) and with small (SAXS) and wide angle X-ray (WAXS) scattering techniques.[5,24] While two dimensional projection TEM images, shown in figure 1,and 2 for $Co_x(SiO_2)_{1-x}$,and $Ni_x(SiO_2)_{1-x}$ respectively,[25] identify these systems as made of metallic nanoparticles, a three dimensional SAXS, probing a larger number of particles, suggests features typical of polydispersive systems. We conclude that the sputtering process generates a variety of amorphous larger structures, nanoparticles, and isolated atoms, regardless the metal transition volume fraction, with about 29 % truly nanoparticles with a volume distribution histogram centered at ~2.3 nm.[24]

**Experimental and spectral analysis**

Temperature dependent near (NIR), medium (MIR), and far infrared (FIR) reflectivity spectra between 30-11000 $cm^{-1}$ were measured in a FT-IR Bruker 113v and a FT-IR Bruker 66 interferometer with 2 $cm^{-1}$ (FIR and MIR) and 6 $cm^{-1}$ (NIR) resolution. Previously to each measurement as prepared films were gently treated by passing on them a lens paper embedded in absolute alcohol. A gold mirror was used as 100% reference.

Our samples were mounted on a cold finger of an Oxford DN 1754 cryostat for measurements between 77 K and 500 K. For lower temperatures we also used a He-close cycle refrigeration system (Displex) coupled to the Bruker 66 reflectivity attachment.

Polarized room temperature angle dependent reflectivity measurements between 20º and 70º have been done using a Spectra Tech Series 500 specular reflectance accessory inserted in the infrared beam path. The radiation was polarized with a polarizer made of a gold wire grid on a KRS-5 substrate.

We estimated phonon frequencies using a standard multioscillator dielectric simulation fit of reflectivity spectra.[26] The dielectric function, $\varepsilon(\omega)$, is given by

$$\varepsilon(\omega) = \varepsilon_1(\omega) + i\varepsilon_2(\omega) = \varepsilon_\infty \prod_j \frac{(\Omega_{jLO}^2 - \omega^2 + i\gamma_{jLO}\omega)}{(\Omega_{jTO}^2 - \omega^2 + i\gamma_{jTO}\omega)} \quad . \qquad (1)$$

We then optimized the calculated reflectivity against the experimental points and thus estimated the high frequency dielectric function, $\varepsilon_\infty$, the transverse and longitudinal optical frequencies, $\Omega_{jTO}$ and $\Omega_{jLO}$, and their transverse and longitudinal damping constants, $\gamma_{jTO}$ and $\gamma_{jLO}$, respectively. We also calculated the Sj strength of the j$^{th}$ oscillator as

$$S_j = \Omega_{jTO}^{-2} \frac{(\prod_k \Omega_{kLO}^2 - \Omega_{jTO}^2)}{(\prod_{k \neq j} \Omega_{kTO}^2 - \Omega_{jTO}^2)} \qquad (2)$$

In addition, when the spectra required it, we added the plasma contribution (Drude term) to the dielectric simulation function as

$$-\frac{(\Omega_{pl}^2 + i(\gamma_{pl} - \gamma_0)\omega)}{(\omega(\omega - i\gamma_0))} \quad , \qquad (3)$$

where $\Omega_{pl}$ is the plasma frequency, $\gamma_{pl}$ its damping, and $\gamma_0$ is understood as a phenomenological damping introduced to reflect lattice drag effects. When these two dampings are set equal, one retrieves the classical Drude formula.[27]

From these analyses the real and imaginary parts of the dielectric function are extracted and used in the calculation of relaxation times as described below. They also allow to calculate the real part of the optical conductivity given by $\sigma_1(\omega) = (\omega/4\pi)\varepsilon_2$ , being $\varepsilon_2$ the imaginary part of the measured dielectric function.

The classical Drude model, describing free carriers contribution centered at zero frequency, may be extended to systems where strong electron-phonon interactions are conspicuous[28,29] by making the damping term complex [30]

This is the so-called extended Drude model[31] useful in describing the coupling any bosonic field to a Fermi liquid. From this, the frequency dependent electronic scattering rate due to many-body interactions at a temperature T, $\left(1/\tau(\omega,T)\right)$, may be written[30]

$$\frac{1}{\tau(\omega)} = \frac{\Omega_{pl}^2}{4\pi} Re\left(\frac{1}{\sigma(\omega)}\right) = \left(\frac{\Omega_{pl}^2}{\omega}\right)\cdot\left(\frac{\varepsilon_2}{(\varepsilon_1^2 + \varepsilon_2^2 + 1 - 2\varepsilon_1)}\right) \qquad (4)$$

Using this formalism, our findings for the three kind of as-deposited films will be presented. The first one as example of those with metal transition volume fractions well above the percolation threshold; a second set of samples with metallic fractions close to metal-insulating transition (where giant enhancements in the ordinary and extraordinary Hall Effect are found); and a third group of samples with metallic concentrations characteristic of insulating films.

## a) $Co_{0.85}(SiO_2)_{0.15}$-a film with metal fraction well above the metal–insulator transition

Figure 3 shows the temperature dependent specular reflectivity of $Co_{0.85}(SiO_2)_{0.15}$ having similar features, as encountered in the current literature for conducting oxides, i.e., as consequence of the transition metal hybridization in the insulating matrix it behaves as a "dirty" metal. That is, from 30 $cm^{-1}$ to 11000 $cm^{-1}$ and from 30 K to 495 K, the spectra have a Drude component, phonons mostly carrier screened at far infrared frequencies, and a long tail that extents toward near infrared. This last associated with hopping electron conductivity and the presence of strong electron-phonon interactions. The temperature dependence of these spectra is representative of any other film in which the transition metal volume fraction implies metallic electron conductivity.

On cooling, the spectral weigh is shifted as the phonon-electron scattering diminishes allowing an increment in the Drude component.

It should also be noted that in order to obtain an excellent tail fit at mid-infrared frequencies it is necessary to introduce highly overdamped oscillator above 1 eV. It may be interpreted as due to photoexcited discrete electronic transitions to mid–gap empty, glass related, defect levels. This coexists with the overwhelming contribution of freer electron hopping that screens this feature. This point will become clear discussing $Ni_{0.61}(SiO_2)_{0.39}$ in the next section.

Figure 4 shows the experimental scattering rate calculated using in equation 4. These results, linear above phonon frequencies, and with strong temperature dependence, is typical of a single kind of carrier (i.e., one relaxation time) under the presence of a strong electron-phonon interactions.

Then, drawn by the remarkable similarity of our spectra with the known for bulk conducting oxides[32], and in an effort to identify the main optical phonon groups involved in the electron-phonon interactions, we have analyzed the optical conductivity, $\sigma_1(\omega)= (\omega/4\pi)\varepsilon_2$ ($\varepsilon_2$ is the imaginary part of the dielectric function), within a small polaron context.

For this purpose, we use the theoretical formulation for small polarons due to nondiagonal phonon transitions as proposed by Reik and Heese[33,34] In this model, optical properties are due to carriers in one small band and interband transitions are excluded. Starting with a Holstein´s Hamiltonian[35] the frequency dependent conductivity is calculated using Kubo´s formula.[36] Then, the real part of the optical conductivity for finite temperature, $\sigma_1(\omega,\beta)$ is given by

$$\sigma_1(\omega,\beta)=\sigma_{DC}\frac{sinh\left(\frac{1}{2}\hbar\omega\beta\right)\exp\left[-\omega^2\tau^2 r(\omega)\right]}{\frac{1}{2}\hbar\omega\beta\left[1+(\omega\tau\Delta)^2\right]^{1/4}}, \qquad (5)$$

$$r(\omega)=\left(\frac{2}{\omega\tau\Delta}\right)\ln\left\{\omega\tau\Delta+\left[1+(\omega\tau\Delta)^2\right]^{1/2}\right\}-\left[\frac{2}{(\omega\tau\Delta)^2}\right]\left\{\left[1+(\omega\tau\Delta)^2\right]^{1/2}-1\right\}, \quad (6)$$

with $\Delta = 2\varpi\tau$ (7)

and $\tau^2 = \frac{\left[ sinh\left( \frac{1}{2}\hbar\varpi\beta \right) \right]}{2\varpi^2\eta}$ . (8)

Here, the conductivity real part, $\sigma_1(\omega,\beta)$, $\beta=1/kT$, is mainly three parameter dependent; $\sigma_{DC}=\sigma(0,\beta)$, the electrical DC conductivity; the frequency $\varpi_j$ that corresponds to the average between the transverse and the longitudinal optical mode of the $j^{th}$ restrahlen band; and $\eta$, a parameter characterizing the strength of the electron-phonon interaction, i.e., the average number of phonons that contribute to the polarization around a localized polaron. It is important to stress that from all these parameters, $\eta$, is the only actually free in the optical conductivity fit for each phonon frequency $\varpi_j$. This is because phonon frequencies are fixed by the reflectivity measurements and the DC-zero frequency-conductivity is known from independent transport measurements.[37] $\eta$ ~3 implies a low to mild electron-phonon interaction while a value around 14 would be corresponding to the very strong end.[38]

The experimental optical conductivities were fitted using that outline. We only assume that the sample frequency dependent conductivity is the convolution of gaussian-like individual contributions each calculated at the phonon frequency $\varpi_j$. It is to note that in this process the resistivity was constrained to positive values. The fit results, applied to every experimental conductivity at a temperature T, are shown in Fig. 5. They show an excellent match with the experimental data and their temperature dependence.

Starting from the lowest measured temperature, at 30 K, (Fig. 5) a good reproduction requires to only allow as variables the first-order main vibrational frequencies of glass $SiO_2$. In spite of omitting an explicit correlation between the different contributions the fitted frequencies are in agreement with the experimental ones, and, as it is intuitively expected, at 30 K the electron-phonon parameters $\eta_j$ (j=1,2,3,4,5), shown in table I, are in a regime for stronger localization. It is worth mentioning that these values for the $\eta$'s, and the individualization of only fundamental lattice frequencies for the low temperature conductivities (highest localization), are findings shared with earlier results in complex oxides.[39] They suggest a low temperature property for the infrared optical conductivities that may apply to all alike oxides in a low temperature regime.

In contrast, when the temperature is raised to 300 K (Fig. 5), the highest frequency vibration mode at 980 $cm^{-1}$, and its overtone 1890 $cm^{-1}$, are singular to the fit at mid-infrared frequencies.[40] That fundamental vibration band, ~980 $cm^{-1}$ in $SiO_2$ , is assigned to the transverse optical of the asymmetric stretching modes (As1,As2) of the Si-O-Si [41,42] . It has the total polarization enhanced by ion motion linked to that vibration.[43]

The detected overtone implies stronger dipole moments generated by polaronic coupling between the charge and the vibrating lattice.[32] Sum phonon processes of different orders are now expected to also contribute to the optical absorption blurring higher assignments due to a number of multiphonon individual combinations.[37] We note that $Co_{0.85}(SiO_2)_{0.15}$ having, at 300 K, $\eta$ 's from 5 to 7 (table I), suggests phonon interactions in the mid to higher strength regime.

## b) $Ni_{0.61}(SiO_2)_{0.39}$- a film near the percolation threshold.

$Ni_{0.61}(SiO_2)_{0.39}$ represents an intermediate case that it is of interest because at about these metal transition concentrations, near the percolation threshold, there are giant enhancements in the two contributions to the Hall resistivity. The ordinary that appears as consequence of the Lorentz force, and the extraordinary one, related to the spin-orbit coupling. We have particularly chosen $Ni_{0.61}(SiO_2)_{0.39}$ because this film also experiences a clear metal to non-metal transition at ~77 K (inset, Fig. 6).

The temperature dependent near normal reflectivity of $Ni_{0.61}(SiO_2)_{0.39}$ (Fig. 6) shows a system in which the reduction in the number of carriers allows less screened phonon bands below 1000 $cm^{-1}$. These appear on the top of a continuum extending into the mid-infrared. At 30 K, in the insulating regime, we did not consider the reflectivity in the higher frequency range (dotted line in Fig. 6) because the film and substrate interface seem to be inducing an interference pattern as the number of carriers is not enough to screen out the substrate.

From them, two main features are distinct in the spectra. One corresponds to a weaker Drude Lorentzian, centered at zero frequency, and another to a wide overdamped oscillator at mid-infrared frequencies. This last is a common feature in oxides that it is usually interpreted in terms of d-d transitions, doped semiconductor, charge density wave ordering, or polarons.[32] Plotted in figure 7, it is apparent that the mid-infrared dipole weakens approaching the metal to insulator transition at ~77 K suggesting a complex carrier behavior in a thermally driven processes. Their scattering rates, using equation 4, yield a clear departure from the linear behavior found in the previous section. Here, as it is shown in figure 8, the curve as close to having a quadratic frequency dependence.

Significantly, Nagel and Schnatterly[44,45] have shown that a two carrier response may be described by a dielectric function as given by

$$\varepsilon(\omega) = 1 - \frac{4\pi N_a e^2}{m_a^*\left(\omega^2 + i\,\omega/\tau_a\right)} - \frac{4\pi N_b e^2}{m_b^*\left(\omega^2 + i\,\omega/\tau_b\right)} \qquad (9)$$

This yields an effective relaxation time

$$\left(\frac{1}{\tau}\right)_{eff} = \frac{1}{\tau_a}\left[1 + \frac{B}{A}\left(\frac{\omega^2 + \tau_a^{-2}}{\omega^2 + \tau_b^{-2}}\right)\right]^{-1} + \frac{1}{\tau_b}\left[1 + \frac{A}{B}\left(\frac{\omega^2 + \tau_b^{-2}}{\omega^2 + \tau_a^{-2}}\right)\right]^{-1} \qquad (10)$$

Then, following references 44 and 45, with A== $4\pi e^2 N_a / m_a^*$,and B= $4\pi e^2 N_b / m_b^*$, and if $\omega\tau_b \ll 1$ or if $\omega\tau_a > 1$, $\omega\tau_b < 1$ and (2B/A) $\omega^2\tau^2_b \ll 1$, equation 10 reduces to

$$(1/\tau)_{eff} \sim (1/\tau)_a + (\omega_{Pb}/\omega_{Pa})\tau_b\omega^2 = a + b\cdot\omega^2 \qquad (11)$$

This, originally developed for noble metal disordered films[44] has the experimental form of $(1/\tau)_{eff}$ found for electron scattering rate of $Ni_{0.61}(SiO_2)_{0.39}$ (fig. 8a). We may then state that the two contributions, yielding a quadratic effective relaxation time – departing from the linear behavior found for $Co_{0.85}(SiO_2)_{0.15}$ (fig. 3)- may be identified as for a sample where two kind of carriers coexist in different environments. In our case, it is likely to be more complex because the glassy environment in which electrons move. Electrons with different electron densities, effective masses, and relaxation times, may be identified as globally belonging to two carrier assemblies[43] in the unavoidable structural imperfections of our samples. The mid-infrared oscillator band (Fig. 7) may then be thought as a profile of convoluted states. In a band picture, they would originate in transitions from valence band to mid-gap empty states, here likely associated to localized electrons in topological glass defects such as dangling bonds and voids. These last playing the role of unintended doping in an amorphous environment.

It was anticipated above the coexistence of localized and itinerant carriers, promoted by the strong interactions between charge carriers and the ions, yields the formation of polarons. The increment in the surface to bulk ratio in the nanoparticles and that transition metal ions in the nanoparticle surface might be uncoordinated (metal nanoaggregatess might obey a dense random packing ordering as in amorphous metals) [46,47] favor an atom relaxation mechanism interacting strongly with the highly polarizable 2p oxygen orbital.

We might, alternatively, have argued that localized self-trapped charges that may be infrared photoexcited are the ones creating the mid-infrared band. However, we feel that the ideas as presented in the preceding paragraphs are better in pointing to glass defects playing a role in the overall mechanism for hopping conductivity in our disordered films.

The underlied polaronic nature of charge carriers in $Ni_{0.61}(SiO_2)_{0.39}$ is supported by the fitting to the experimental conductivities using equation 6. The fit parameters are shown in table II. The temperature dependent optical conductivities of $Ni_{0.61}(SiO_2)_{0.39}$, Fig. 9, show results similar to those for $Co_{0.85}(SiO_2)_{0.15}$. The η´s (table II) characterizing the strength of the electron-phonon interaction suggests stronger localization, and, also show that as we increase the temperature higher order phonons play non –negligible role at mid-infrared frequencies.

We also note that the consequences of electron thermal localization may quantitatively be estimated using the sum rule that allows to calculate the area ratio of the mid-infrared optical conductivity at 495 K and 77K. Thus,

$$N_{eff}(\omega)=\frac{2mV_{cell}}{\pi e^2}\int_0^{11000}\sigma_1(\omega')\cdot d\omega' \qquad (12)$$

where $N_{eff}$ is an effective carrier number; $V_{cell}$ a "volume cell"; m, and e are the mass and charge, respectively, yields a rough estimated of about 82% on the relative number of localized carriers when the film cools to 77 K.

## c) $TM_x(SiO_2)_{1-x}$ 0.55 > x > 0.0, TM = Co, Ni, Fe- films with metal fractions below the metal-insulator transition.

We will now comment on our results for the case in which the amount of transition metal in the films is such that most of the conducting critical paths are truncated. Thus, as it is shown in Fig. 10 for $Co_{0.51}(SiO_2)_{0.49}$, the film near normal specular reflectivity displays well defined vibrational bands and a sharp threshold at ~1250 $cm^{-1}$ (we ignore features at frequencies higher than ~2100 $cm^{-1}$ because at about that value $SiO_2$ has the transmission edge implying a wavy interference pattern distorting the spectrum at higher frequencies). It also shows that all bands undergo a red-shift relative to pure $SiO_2$ glass (dashed line in Fig. 10). This kind of frequency shifts have been already found in ion implanted and irradiated glass samples where the bond angle tend to be smaller.[48] When the average Si-O-Si bond angle is reduced the fundamental Si-O asymmetric stretching vibration at ~1000 $cm^{-1}$ shifts down in frequency.[49] We understand that shift as due to topological errors result of introducing the transition metal ions in the glass network with a possible contribution of strong electron-phonon interactions and localization.

We also note an unlike behavior for the reflectivity minimum in the carrier-longitudinal optical mode (LO) mode interaction at 1210 $cm^{-1}$.

It is known that on cooling the reflectivity minimum, at approximately the highest LO frequency, undergoes when there are strong electron-phonon interactions a red shift relative to other phonon features.[50] Moreover the band centered at ~1450 $cm^{-1}$, assigned to the electron promotion to higher energy available levels, is regularly found to increase in intensity. (e.g., see for $Sr_2FeMo_{0.2}W_{0.8}O_6$ in ref. 51 (inset, Fig. 7)). Contrasting, the band minimum for $Co_{0.51}(SiO_2)_{0.49}$ at 1210 $cm^{-1}$ (Fig. 10) although reproduces the behavior expected for strong electron-phonon interactions does also suggests that no much else is taking place in a wide range of temperatures.

For this reason, to search and enhance any longitudinal contribution that might be hiding in the spectra, we also performed angle dependent reflectivity at room temperature.

Oblique reflectivity has been regularly use for detecting longitudinal optical modes, lattice vibrations perpendicular to a film surface, in transmission mode. Berreman[52] first pointed out that having radiation polarized parallel and perpendicular to the plane defined by the incident radiation and the normal to the film surface would allow to directly measure longitudinal optical modes. Since transverse magnetic (TM) P-polarized radiation has a wave vector with components parallel and perpendicular to the film surface extra bands not seen in transverse electric S-polarization will appear.

In normal incidence, longitunal modes are found at the minima (zero) of the dielectric function or a maximum of $\mathrm{Im}\left\{\frac{-1}{\varepsilon}\right\}$. In oblique reflectivity, the energy loss for incident radiation transverse magnetic polarized reduces to

$$P \approx \frac{\varepsilon_0 \omega}{2} \mathrm{Im}\{\varepsilon\} \cdot |E_N|^2 = \frac{\mu_0 \omega}{2} \sin^2 \alpha \cdot \mathrm{Im}\left\{\frac{-1}{\varepsilon}\right\} \cdot |H|^2 \qquad (13)$$

where $E_N$ is the electric normal field, H the magnetic field parallel to the interfaces and $\alpha$ is the angle of incidence. Thus, at the longitudinal modes the spectra will have a distinctive enhancement due to the maximum of $\mathrm{Im}\left\{\frac{-1}{\varepsilon}\right\}$ in contrast to transverse optical modes intensities.[53]

Using this principle on glasses Almeida[54] showed that glass-$SiO_2$ sustains clear TO-LO splits for the main vibrational bands. Reproduced by us in Fig. 11, the profiles show that by increasing the angle of incidence the overall band weigh shifts to higher frequencies peaking as expected at longitudinal modes. Main bands are rather broad because for each glass vibrational splitting there might be several like-modes contributing to the one band profile. We will call them Berreman phonons to differentiate those structures from an unexpected feature that becomes active in the reflectivity of our films at about the same frequencies.

Fig. 12 shows the P-polarized angle dependent specular reflectivity of a number of insulating granular films. We note that as the angle increases in addition to much ill defined Berreman longitudinal phonons a sharp peak appears at the very edge of the 1450 $cm^{-1}$ band. As it is seen for $Co_{0.38}(SiO_2)_{0.62}$, $Fe_{0.34}(SiO_2)_{0.66}$, $Ni_{0.28}(SiO_2)_{0.72}$. this peak becomes stronger with increasing angle and at ~70 º is the dominant feature.

To our knowledge, there is only one report that discusses a comparable phenomenon. Ahn et al[55] working on $Fe_3O_4$ films deposited on a MgO substrate found a resonant band emerging on the top of the film reflectivity but related to the MgO LO phonons. We believe that the appearance of this resonance in our films has the same physical origin. Nominally, replacing the conducting $Fe_3O_4$ films by a metal nanoparticle and insulator MgO by the $SiO_2$ glass matrix we may understand our physical picture using the boundary conditions as proposed in reference 54.

The normal ($E_N$) and tangential ($E_T$) component at the boundary for the incident to the film electric field are then given by[53]

$$E_N = -\sin\alpha \cdot \sqrt{\mu_0 / \varepsilon_0} \cdot H \cdot \begin{cases} 1/\varepsilon & \text{On the metal side} \\ 1/\varepsilon_{matrix} & \text{On the glass side} \end{cases} \qquad (14)$$

$$E_T = \sqrt{\varepsilon_{matrix} - \sin^2\alpha} \cdot \sqrt{\mu_0 / \varepsilon_0} \cdot H \cdot 1/\varepsilon_{matrix} \qquad \text{On both sides} \quad (15)$$

We see that the tangential component $E_T$ has a null condition at $\varepsilon(\omega) \sim \sin^2\alpha$. When this is satisfied, the field incident has only the normal component that in turn excites electrons in the nanoparticle. Thus, once the appropriate angle is reached, a collective electron oscillation is induced as a localized plasma with origin in the metallic nanostructure. The resonant plasma band, as it is shown in Fig. 12, appears at or slighted lower energy of the edge denoting the electron promotion by energetic enough photons. We then reason that the origin of the collective electron oscillation may be traced to electrons that are not able to overcome the metal-dielectric interface of the transition metal nanoparticles embedded in $SiO_2$. Being localized, the collective cloud pairs with the positive ion background to yield a strong infrared active electric dipole. The coupled oscillation -wave propagating photon-electron cloud- is known as plasma polariton.[56] On the other hand, the role of the nanoparticle irregular surface remains to be established since at metal-dielectric interfaces electron oscillations are confined to the surface and localization leads to giant enhancement of the local and electrical fields. In surface enhanced Raman scattering (SERS) this principle. i.e, plasma localization in rough surfaces, is used to resolve molecule vibrational modes with highest resolution.[57]

The longitudinal vibrations of $Fe_{0.34}(SiO_2)_{0.66}$, as it was shown in Fig. 11 for $SiO_2$, have stronger bands and shift in frequency as expected for the LO mode following eq. 13 in every reststrahlen band. As the reflectivity angle is increased to 70º (Fig. 13b) the longitudinal mode merges with the resonant band inducing broadening and softening of the overall band profile.

The longitudinal character of the resonant electronic cloud may be visualized in Fig. 13 where the S-(TE) and P-(TM) polarized spectra for $Fe_{0.34}(SiO_2)_{0.66}$ are plotted. Fig. 13a shows for 50º incidence the dissimilar profiles of the vibrational band, assigned to TO-LO antisymmetric stretching, against that of the resonance. This last has a corresponding S-polarized sharp antiresonance confirming its purely longitudinal character even for intermediate angles of incidence. When we plot (Fig. 14) peak positions against the incident angle we find two regimes. At low angles, the almost linear behavior is associated with the resonant condition leading the infrared response while a strong departure from it is found at higher angles. It is as if the glass LO modes, interacting with the nanoplasma, experience a frequency shift reminiscent to phonons in oxides in an environment with strong electron-phonon interactions.[49,50]

## Conclusions

Summarizing, we have shown the infrared response for transition metal nanograins embedded in $SiO_2$ for three specific metal fractions. Well above the percolation threshold the films, regardless the transition metal, have the behavior known for conducting oxides with the characteristic hopping conductivity as transport mechanism. At intermediate concentrations the main feature in the infrared spectra, in addition to a Drude term, is a mid-infrared band that in the case of $Ni_{0.61}(SiO_2)_{0.39}$ follows a thermally driven metal-non metal phase transition. The polaron fit to the optical conductivity here shows a localization and stronger electron-phonon interaction.

A remarkable nanoplasma resonance is found in globally insulating films strongly interacting with Berreman phonons. Common to all our films the key for its unambiguous detection is the reduced number of free carriers. As it is as shown in figure 15 for $Ni_{0.61}(SiO_2)_{0.39}$, if the film have enough hopping electrons the resonant peak will be only detected at the highest oblique configurations and even then, it would easily be confused by a regular phonon feature.

It is important to emphasize that hopping conductivity-polaron formation-; mid gap band- assimilated to an overdamped oscillator-, and the resonant condition are features always present for all metal fractions with different degree of occurrence. It depends on the fractional amount of transitional metal nanoparticles, and thus, on the number of carriers. The evolution of transport properties relies on the behavior of insulating and metallic aggregates reminiscent of highly correlated systems where there is coexistence of competing states in an inhomogeneous environment with nanometric scaling.[58,59,60,61]

We may then conclude that the nanoplasma resonant band addressed in the above sections may turn to be an useful tool to optically monitor with great detail a metal to insulator transition. While the resonance will be screened in a sample with enough hopping electrons it will be a sharp feature in globally insulating environments.

**Acknowledgements**

This work was partially financed by the Foundation for Harboring Research of the State of Sao Paulo (Fundação de Amparo dà Pesquisa do Estado de São Paulo-FAPESP) and by the National Council for Technological and Scientific Development (Conselho Nacional de Desenvolvimento Cientìfico e Tecnológico-CNPq), Brazil. J.C.D. is grateful to the LatinAmerican Center for Physics (Centro Latinoamericano de Física-CLAF) for a travel grant. X.X.Z. acknowledges financial support from the Hong Kong Special Administrative Region, China (Project No605704). N.E.M. acknowledges financial support from the Argentinean National Research Council (Consejo Nacional de Investigaciones Científicas y Técnicas-CONICET) (Project: PIP 5152/06).

# REFERENCES

1. Tripathi and Y. L. Loh, ArXiv: cond-mat 1138 v1 (2006), and references therein.
2. A.B. Pakhomov, S.K. Wong, X. Yan and X.X. Zhang, Phys. Rev. B **58**, R13375 (1998).
3. X.X. Zhang, C. Wan. H. liu, Z. Q. li, P. Sheng and J.. L. Lin, Phys. Rev. Lett. **86**, 5562 (2001); A. B. Pakhomov, X. Yan and B. Zhao; Appl. Phys. Lett. **76**, 3497 (1995).
4. J. C. Denardin, M. Knobel, X.X. Zhang, A. B. Pakhomov, J. of Magnetism and Magnetic Materials **262**, 15 (2003); J. C. Denardin, A. B. Pakhomov, M- Knobel, H. Liu, and X.X.Zhang, J. Phys. Condens. Matter **12**, 3397 (2000).
5. L.M. Socolovsky, C.P.L. Oliveira, J.C. Denardin, M. Knobel, and I. Torriani, J. of Appl. Physics **99**, 08C511 (2006).
6. For a recent review see, I. S. Beloborodov, A. V. Lopatin, and V. M. Vinokur, and K. B. Efetov, Rev. of Mod. Physics **79** , 469 (2007).
7. T. Vicsek , Nature **418**, 131 (2002).
8. J.Seebeck, T. R. Nielsen, P. Gartner, and F. Jahnke . Phys. Rev. B **71**, 125327 (2005).
9. E. Dagotto, Science **308**, 257 (2005).
10. R. P. Devaty and A. J. Sievers, Phys. Rev. B **41**, 7421 (1990)
11. K. D. Cummings, J. C. Garland, and D. B. Tanner, Phys. Rev B **30**, 4170 (1984)
12. Y. H. Kim and D. B. Tanner, **39**, 3585 (1989).
13. T. W. Noh, Y Song, S. Lee and J. R. Gaines, Phys. Rev. 33, 3793 (1986).

14. Philos. Trans. R. Soc. London **203**, 385 (1904); ibidem **205**, 237 (1906)
15. D. Stroud, Phys. Rev. B **12**, 3368 (1975)
16. D.A. G. Brugggeman Ann. Phys. (Leipzig) **24**, 636 (1935)
17. R. Landauer J. Appl. Phys. **23** , 779 (1952).
18. R. W. Cohen, G. D. Cady, M. D. Cody, M. D. Coutts, and B. Abeles, Phys. Rev. B **8**, 3689 (1973).
19. O. A. Yeshchenko, I. M. Dmitruk, A. A. Alexeenko, and A. M. Dmytruk, Journal of Physics and Chemistry of Solids **69**, 1615 (2008).
20. O. A. Yeshchenko, I. M. Dmitruk, A. A. Alexeenko, and A. M. Dmytruk, Applied Surface Science 254, 2736 (2008)
21. P. Gao, T. A. Tyson, Z. Liu, M. A. DeLeon, and C. Dubourdieu, arXiv:0803.3182.
22. V. Dobrosavljevc´, D. Tanaskovic´ and A. A. Pastor , Phys. Rev. Lett. **90**, 016402 (2003).
23. I. Romero, J. Aizpurua, G. W. Bryant, and F. J. García de Abajo, Opt. Express **14**, 9988 (2006).
24. I. L. M. Socolovsky, C. L. P. Oliveria, J. C. Denardin, M. Knobel, and I. L. Torriani, Phys. Rev. B **72**, 184423 (2005).
25. J.C.Denardin, Ph.D thesis, Instituto de Física, Universidade Estadual de Campinas, Campinas, SP, Brazil.
26. T. Kurosawa, J. Phys. Soc. Jpn **16**, 1208 (1961).
27. F. Gervais, J. L. Servoin, A. Baratoff, J. B. Berdnorz, and G. Binnig, Phys. Rev. **47**, 8187 (1993).
28. P. B. Allen, Phys. Rev. B **3**, 305 (1971).
29. P. B. Allen and J. C. Mikkelsen, Phys. Rev. B **15**, 2952 (1997).
30. A. V. Puchkov, D. N. Basov, and T. Timusk, J. Phys. : Condensed Matter **8**, 10049 (1996).

31. D. N. Basov and T. Timusk, Rev. Mod. Phys. **77**, 721 (2005).
32. P. Calvani. "*Optical Properties of Polarons*", La Rivista del Nuovo Cimento, (8) (2001).
33. H. G. Reik, in *Polarons in Ionic Crystals and Polar Semiconductors*, edited by J. Devreese (North-Holland, Amsterdam), 1972.
34. H. G. Reik and D. Heese, J. Phys. Chem. Solids **28**, 581,(1967).
35. T. Holstein, Ann. Phys. (NY) **8**, 343 (1959).
36. R. Kubo, J. Phys. Soc. Jpn **12**, 570 (1957).
37. R. Mühlstroh and H. G. Reik , Phys. Rev. **162**, 703 (1967).
38. H. G. Reik , Zeitschrift für Physik 203, 346 (1967).
39. N. E. Massa, H. Falcon, H. Salva and R. E. Carbonio, Phys. Rev. B **56**, 10178 (1997).
40. N. E. Massa, J. C. Denardin, L. M. Socolovsky, M. Knobel, F. P. de la Cruz, and X.X. Zhang, Solid State Communications **141**, 4551 (2007).Ibidem, http://meetings.aps.org/link/BAPS.2006.MAR.D15.8
41. P. C. Coombs, J. F. DeNatale, P.J. Hood, D. K. McElfrsh, R. S. Wortman, and J. F. Shackelford, Phys. Mag. B **51** L39 (1985).
42. F. L. Galeener, R. A. Barrio, E. Martinez, and R. J. Elliot, Phys. Rev. Lett. **53**, 2429 (1984).
43. M. Wilson, P. A. Madden, M. Hemmati and C. A. Angeli, Phys. Rev. Lett. **77**, 4023 (1996).
44. R. Nagel and S. E. Schnatterly , Phys Rev. **9**, 1299 (1974).
45. S. R. Nagel and S. E. Schnatterly, Phys. Rev. B **12**, 6002 (1975).
46. L. V. Heimendahl, J. Phys. F. Met. Phys.. **5**, L141 (1975)-
47. P. Kluth, B. Johannessen, G. J. Foran, D. J. Cookson, S. M. Kluth and M. C. Ridgway Phys. Rev. B **74**, 014202 (2006) references therein-
48. I. Simon, J. Am. Ceram. Soc. **40**, 159 (1957).

49. A. Agarwal, K. M. Davis, and M. Tomozawa, J. Non-Cryst. Solids **185**, 191 (1995)-
50. M. Reizer, Phys. Rev. B **61**, 40 (2000).
51. N. E. Massa, J. A. Alonso, M. J. Martinez-Lope and M. T. Casais, Phys. Rev. B **72**, 214303 (2005).
52. D. W. Berreman, Phys. Rev. **130**, 2193 (1963).
53. B. Herbecke, B. Heinz, and P. Grosse, Appl. Phys. A **38**, 263 (1985).
54. R. M. Almeida, Phys. Rev B **45**, 161 (1992).
55. J. S. Ahn, H. S. Choi and T. W. Noh, Phys. Rev. **53**, 10310, (1996).
56. V. V. Klimov and D.V. Guzatov, Appl. Phys. **A89**, 305 (2007).
57. M. Moskovits Rev. Mod. Phys. **57**, 783 (1985).
58. Z. Sun, J. F. Douglas, A. V. Fedorov, Y.-D. Chuang, H. Zheng, J. F. Mitchell, and D. S. Dessau. Nature Physics **3**, 248 (2007).
59. M. M. Qazilbash, M. Brehm, Byung-Gyu Chae, P.-C. Ho, G. O. Andreev, Bong-Jun Kim, Sun Jin Yun, A. V. Balatsky, M. B. Maple, F. Keilmann, Hyun-Tak Kim, and D. N. Basov, Science **318**, 1750 (2007). Ibidem , http://arxiv.org/abs/0801.1171.
60. D. Kumar, K. P. Rajeev, J. A. Alonso, and M. J. Martínez-Lope, ArXiv: [cond.- matter.other]. 0712.3990v1 (2007).
61. T. V. Shubina, S. V. Ivanov, V. N. Jmerik, D.D. Solnyshkov, V. A. Vekshin, P. S. Kop'ev, A. Vasson, J. Leymarie, A. Kavolin, H. Amano, K. Shimono, A. Kasic and B. Monemar, Phys. Rev. Lett. **92**, 117407 (2004).

# Table I

Parameters of the small polaron theory fit for the optical conductivity of $Co_{0.85}(SiO_2)_{0.15}$. Note that vibrational frequencies are in agreement with the film IR experimental bands (in brakets) and that the conductivity infrared tail at higher than 30 K is dominated by the overtone of the vibrational band at ~980 $cm^{-1}$. The fit resistivities, $\rho_{DC}$, are also included.

| T K | $\rho_{DC}$ (ohm-cm) | $\eta_1$ | $\varpi_{ph1}$ ($cm^{-1}$) | $\eta_2$ | $\varpi_{ph2}$ ($cm^{-1}$) | $\eta_3$ | $\varpi_{ph3}$ ($cm^{-1}$) | $\eta_4$ | $\varpi_{ph4}$ ($cm^{-1}$) | $\eta_5$ | $\varpi_{ph5}$ ($cm^{-1}$) |
|---|---|---|---|---|---|---|---|---|---|---|---|
| **30** | 0.000260 | 4.65 | 275.0 (280) | 5.7 | 357.1 (361.3) | 8.02 | 481.3 (400) | 10.81 | 670.7 (575) | 14.4 | 940. (980) |
| 77 | 0.000488 | 2.0 | 281.0 (280) | 3.8 | 385.0 (381) | 5.46 | 599. (575) | 7.4 | 943. (980) | 10.5 | 1469. |
| **300** | 0.002416 | 5.2 | 283.0 | 4.9 | 380. | 5.51 | 560. | 5.72 | 1070. | 7.12 | 1950. |
| **495** | 0.002569 | 9.4 | 280. | 5.9 | 340. | 5.7 | 590. | 5.65 | 1080 | 6.47 | 2040. |

## Table II

Examples of parameters for the small polaron theory fits to the optical conductivity of $Ni_{0.61}(SiO_2)_{0.39}$. Note that vibrational frequencies are in agreement with the film IR experimental bands (in brackets) and that the conductivity tail for temperatures higher than 77 K is dominated by the overtone of the highest vibrational band at ~1000 $cm^{-1}$. The fit resistivities, $\rho_{DC}$, are also included. For 77 K we also added an extra contribution as a low frequency phonon band at 210 $cm^{-1}$ due the resonance (see text) distorting the conductivity edge. This effect was not considered for 200 K.

| T K | $\rho_{DC}$ (ohm-cm) | $\eta_1$ | $\varpi_{ph1}$ ($cm^{-1}$) | $\eta_2$ | $\varpi_{ph2}$ ($cm^{-1}$) | $\eta_3$ | $\varpi_{ph3}$ ($cm^{-1}$) | $\eta_4$ | $\varpi_{ph4}$ ($cm^{-1}$) | $\eta_5$ | $\varpi_{ph5}$ ($cm^{-1}$) | $\eta_6$ | $\varpi_{ph6}$ ($cm^{-1}$) |
|---|---|---|---|---|---|---|---|---|---|---|---|---|---|
| 77 | 0.00765 | 8.19 | 210. | 7.17 | 410. (423) | 9.18 | 483. (495) | 10.5 | 591. (584) | 11.66 | 820- (832) | 7.85 | 1100. (1126) |
| **200** | 0.00770 | 6.9 | 430. (423) | 7.5 | 490. (495) | 7.2 | 571. (584.) | 7.7 | 830. (832) | 8.7 | 1100. (1126) | 11.3 | 1984. |

# FIGURE CAPTIONS

Figure 1.*(color online*) Examples of TEM images for $Co_x(SiO_2)_{1-x}$, a) x=0.28, b) x=0.45, c) x=0.62, b) x=0.77. Inset : X-ray diffraction pattern for $Co_{0.77}(SiO_2)_{0.23}$ (after ref. 25)

Figure 2.*(color online*) Examples of TEM images for $Ni_x(SiO_2)_{1-x}$, a) x=0.30, b) x=0.40, c) x=0.50, b) x=0.61. Inset : X-ray diffraction pattern for $Ni_{0.61}(SiO_2)_{0.39}$ (after ref. 25)

Figure 3.*(color online*) Temperature dependent reflectivity of $Co_{0.85}(SiO_2)_{0.15}$ Inset: Same sample metallic behavior. Note that here we plot resistance instead of resistivity because, although with almost the same slope, we found changes when probing the resistivity at different sites in a same film.

Figure 4.*(color online)* Temperature dependent scattering rate for $Co_{0.85}(SiO_2)_{0.15}$

Figure 5.*(color online)* Temperature dependent optical conductivity of $Co_{0.85}(SiO_2)_{0.15}$ Full line: experimental; superposing dots : fit**.** The fit assumes that each conductivity may be though a sum of gaussian-like (eq. 6) contributions each calculated at a phonon frequency $\varpi_j$ and at a temperature T (see text and table I).

Figure 6.*(color online)* Temperature dependent reflectivity of $Ni_{0.61}(SiO_2)_{0.39}$ . Inset: Same sample resistivity

Figure 7.*(color online)* Temperature dependent Drude and mid-infrared oscillator for $Ni_{0.61}(SiO_2)_{0.39}$. The inset shows the same sample resistivity.

Figure 8.*(color online)* Temperature dependent scattering rate for $Ni_{0.61}(SiO_2)_{0.39}$. a) linear plot; b) Same data in a log-log plot to showing the close quadratic behavior.

Figure 9.*(color online)* Temperature dependent optical conductivity of $Ni_{0.61}(SiO_2)_{0.39}$ Full line: experimental; superposing dot : fit**.** The fit assumes that each conductivity may be though a sum of gaussian-like (eq. 6) contributions each calculated at a phonon frequency $\varpi_j$ and at a temperature T (see text and table I). (see text and table II).

Figure 10.*(color online)*Temperature dependent reflectivity of $Co_{0.51}(SiO_2)_{0.49}$. Dashed line is the $SiO_2$ glass reflectivity at room temperature. Inset: Same sample resistivity.

Figure 11.*(color online)* Oblique infrared reflectivity of $SiO_2$ at different incident angles.

Figure 12.*(color online)* Angle dependent oblique P-polarized infrared reflectivity of $Co_{0.38}(SiO_2)_{0.62}$, $Fe_{0.34}(SiO_2)_{0.66}$ and $Ni_{0.28}(SiO_2)_{0.72}$ at 300 K. The circles enhance the onset of the remarkable nanoplasma resonance at the band threshold.

Figure 13.*(color on line)* 50º and 70º oblique incident $Fe_{0.34}(SiO_2)_{0.66}$ S- and P-reflectivity near the infrared nanoplasma resonant frequency at 300 K .

Figure 14.*(color online)* Nanoplasma resonant peak frequency versus angle of incidence for $Co_{0.38}(SiO_2)_{0.62}$, $Fe_{0.34}(SiO_2)_{0.66}$ and $Ni_{0.28}(SiO_2)_{0.72}$.. Dashed line is a guide for the eye.

Figure 15.*(color online)* Angle dependent oblique P-polarized infrared reflectivity of $Ni_{0.61}(SiO_2)_{0.39}$. at 300 K . The circle points to the detection of the nanoplasma resonance in a electron screened environment .

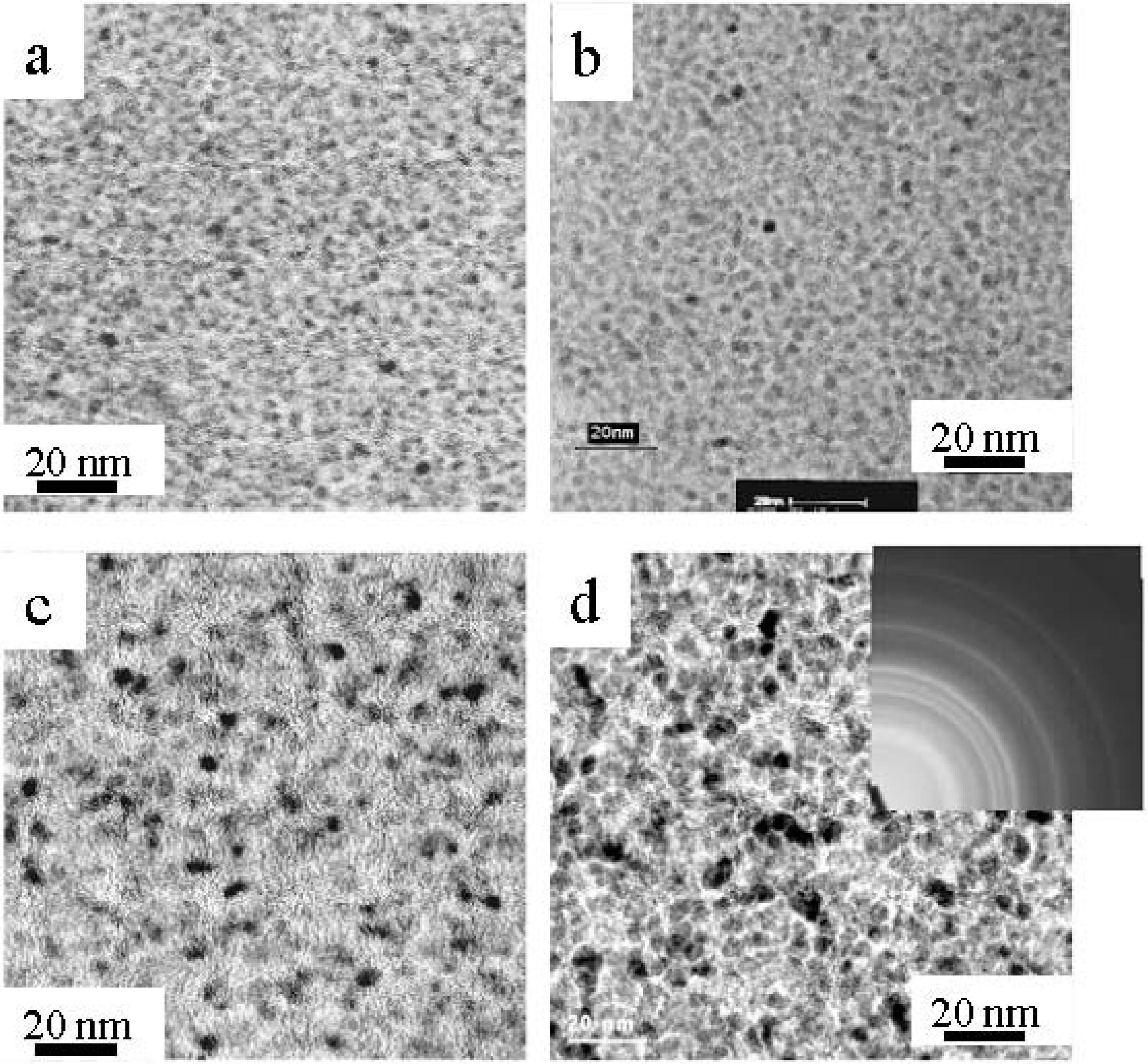

Figure 1
Massa et al

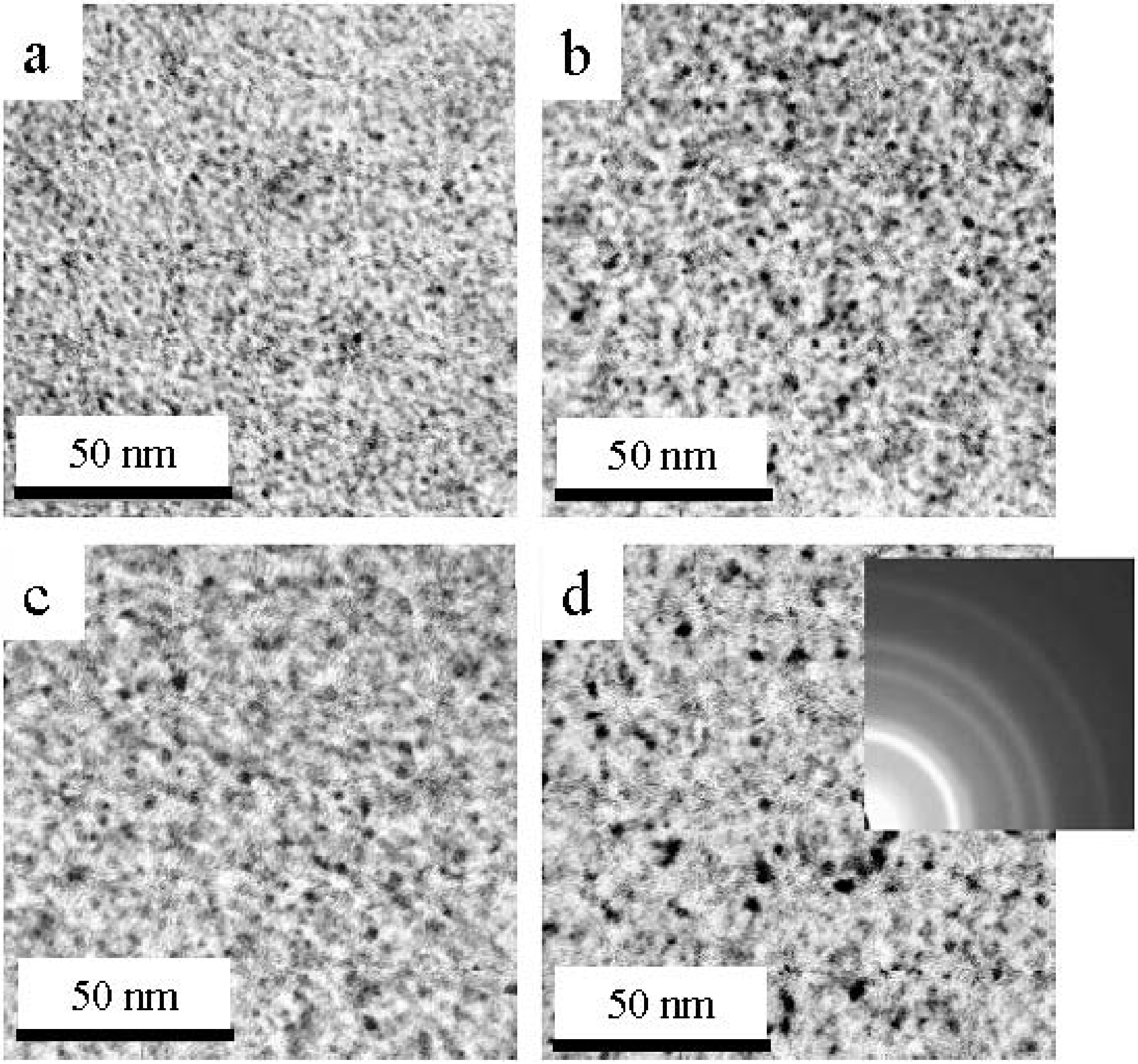

Figure 2
Massa et al

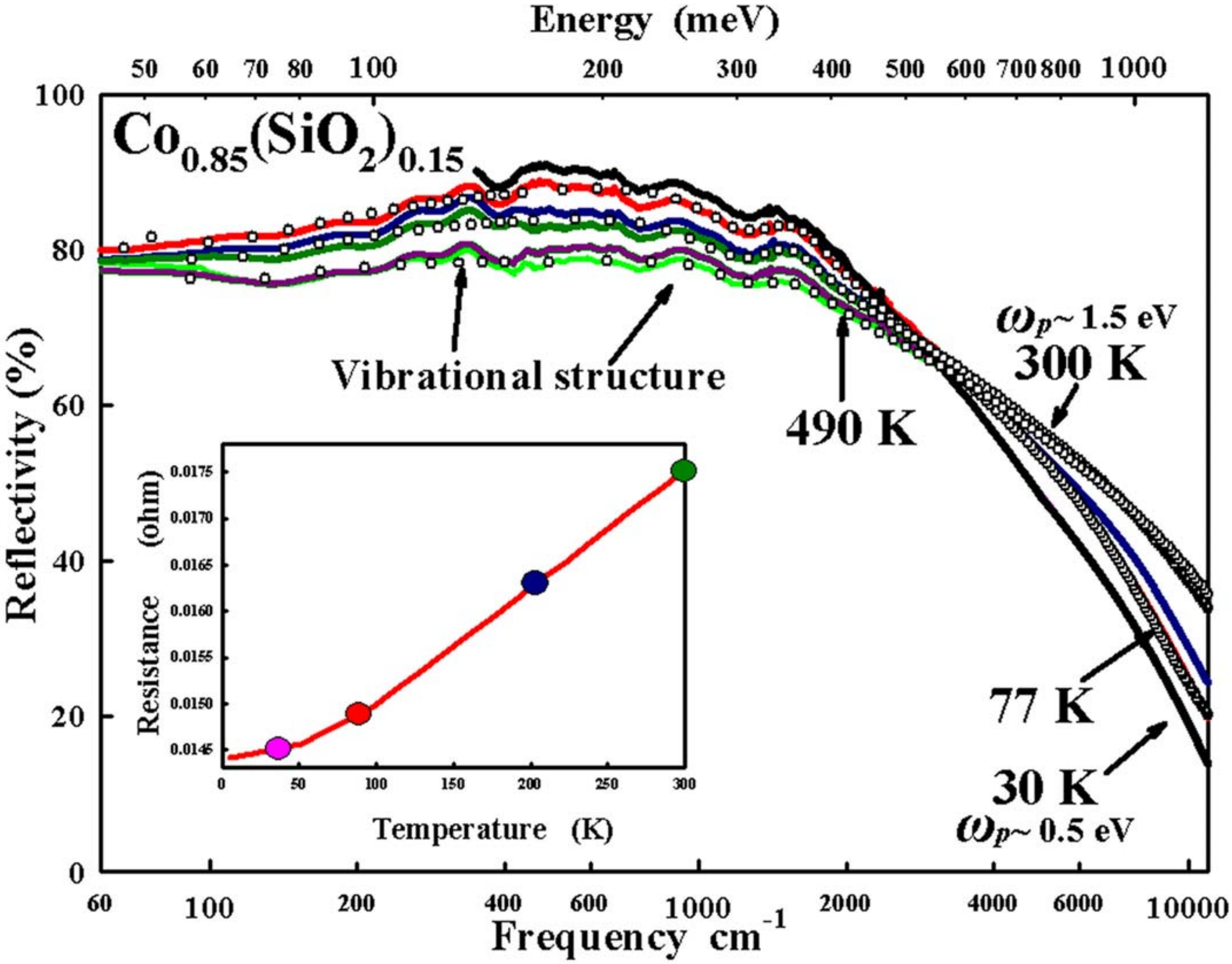

**Figure 3**
**Massa et al**

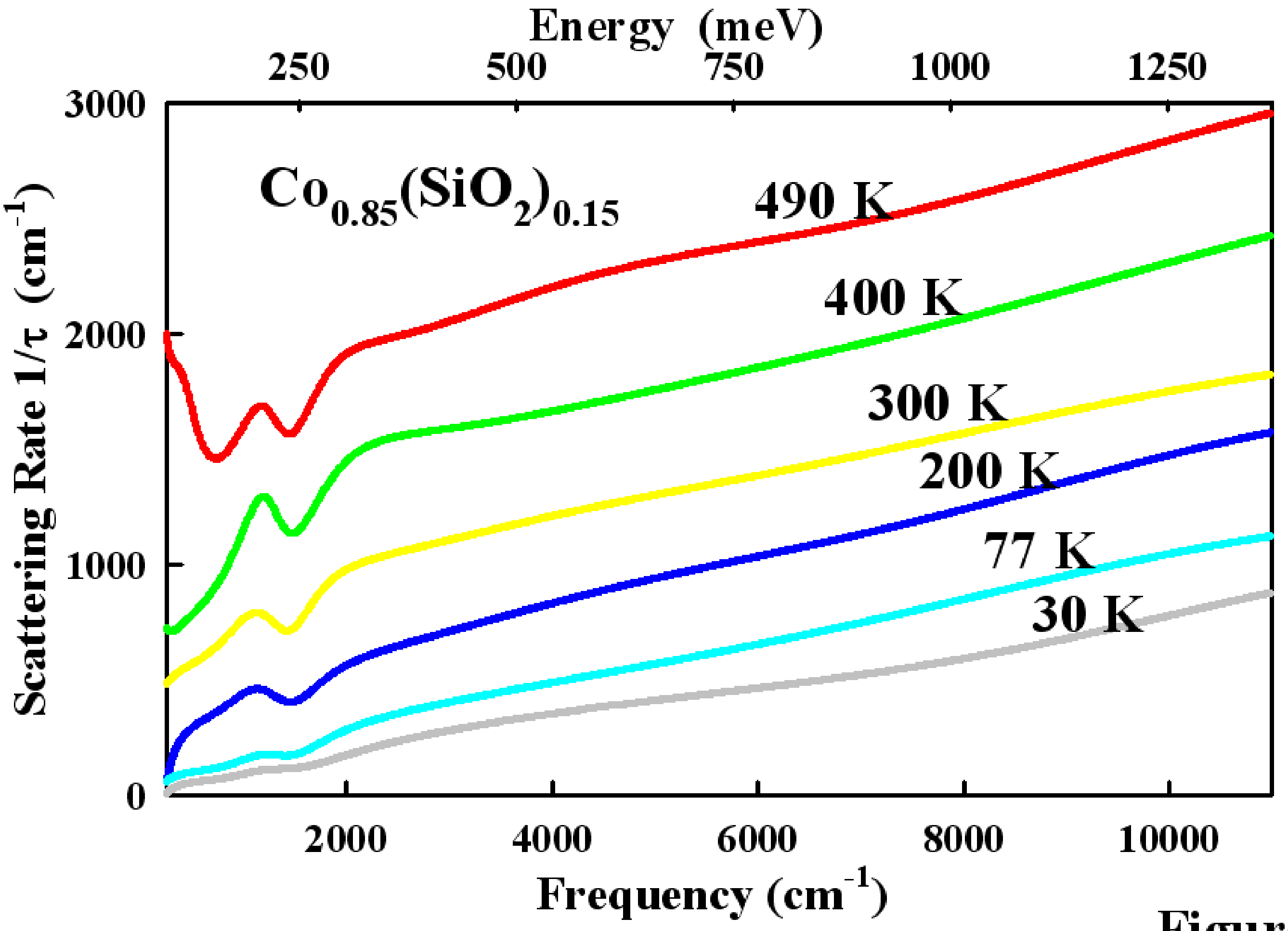

**Figure 4**
**Massa et al**

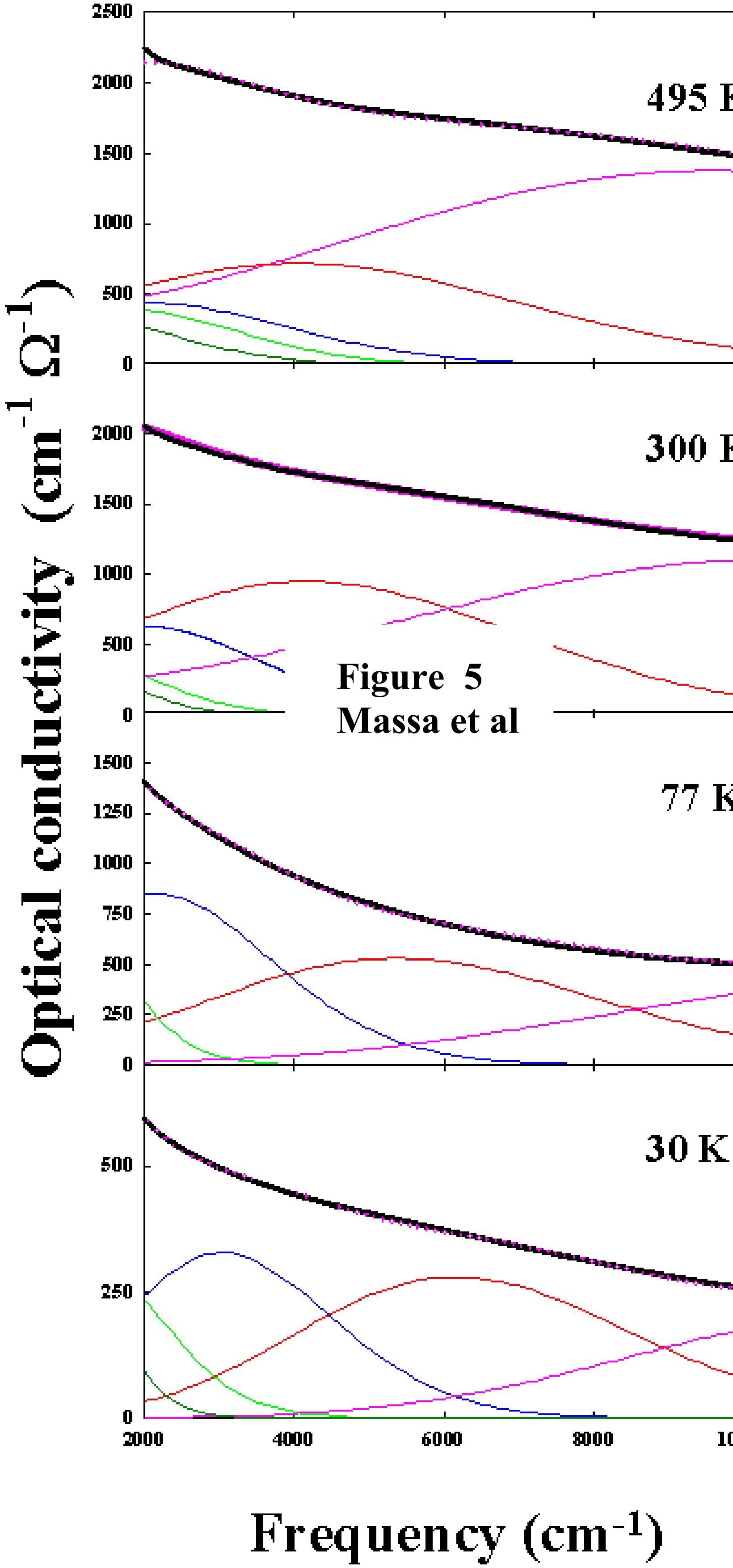

Figure 5
Massa et al

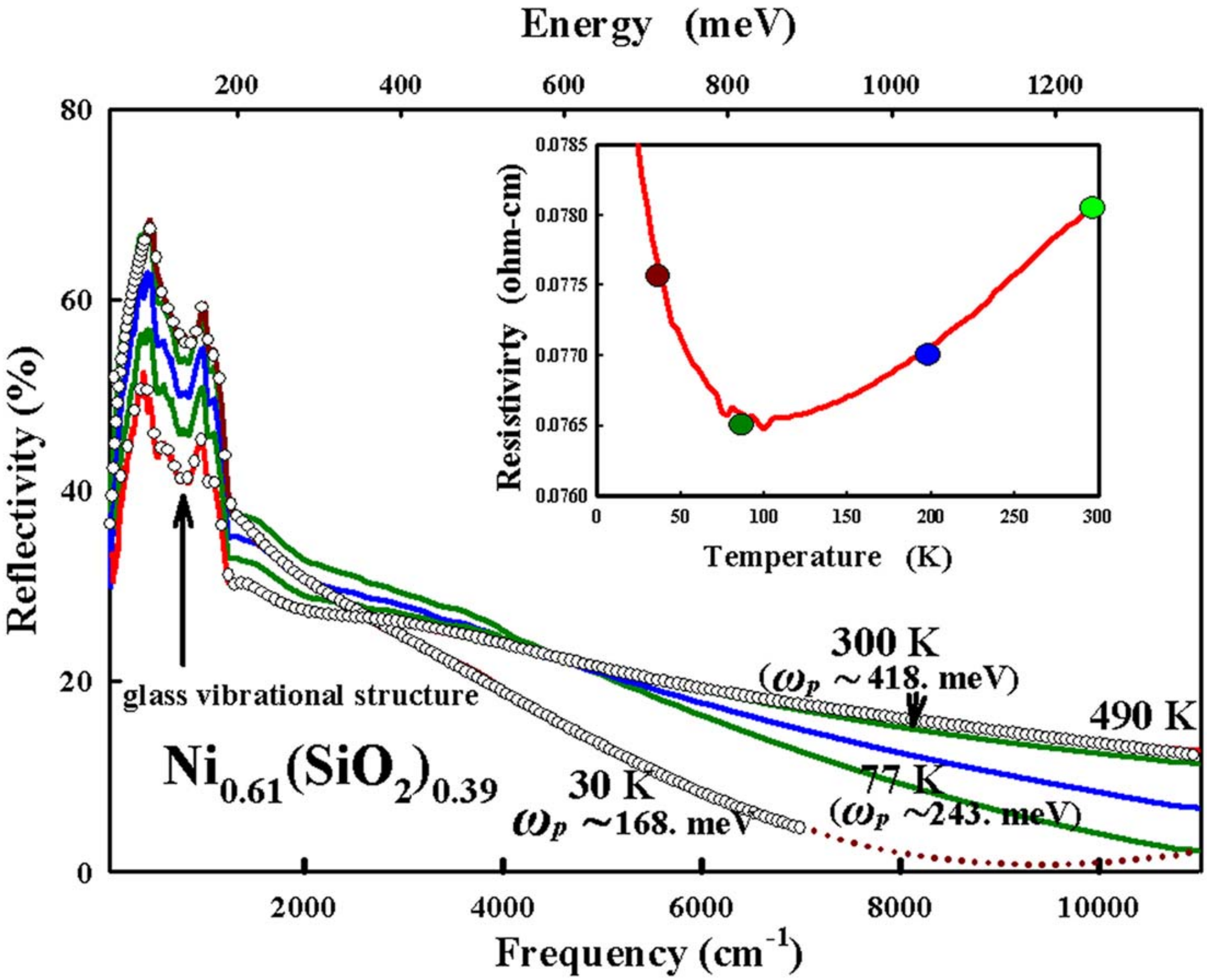

**Figure 6**
**Massa et al**

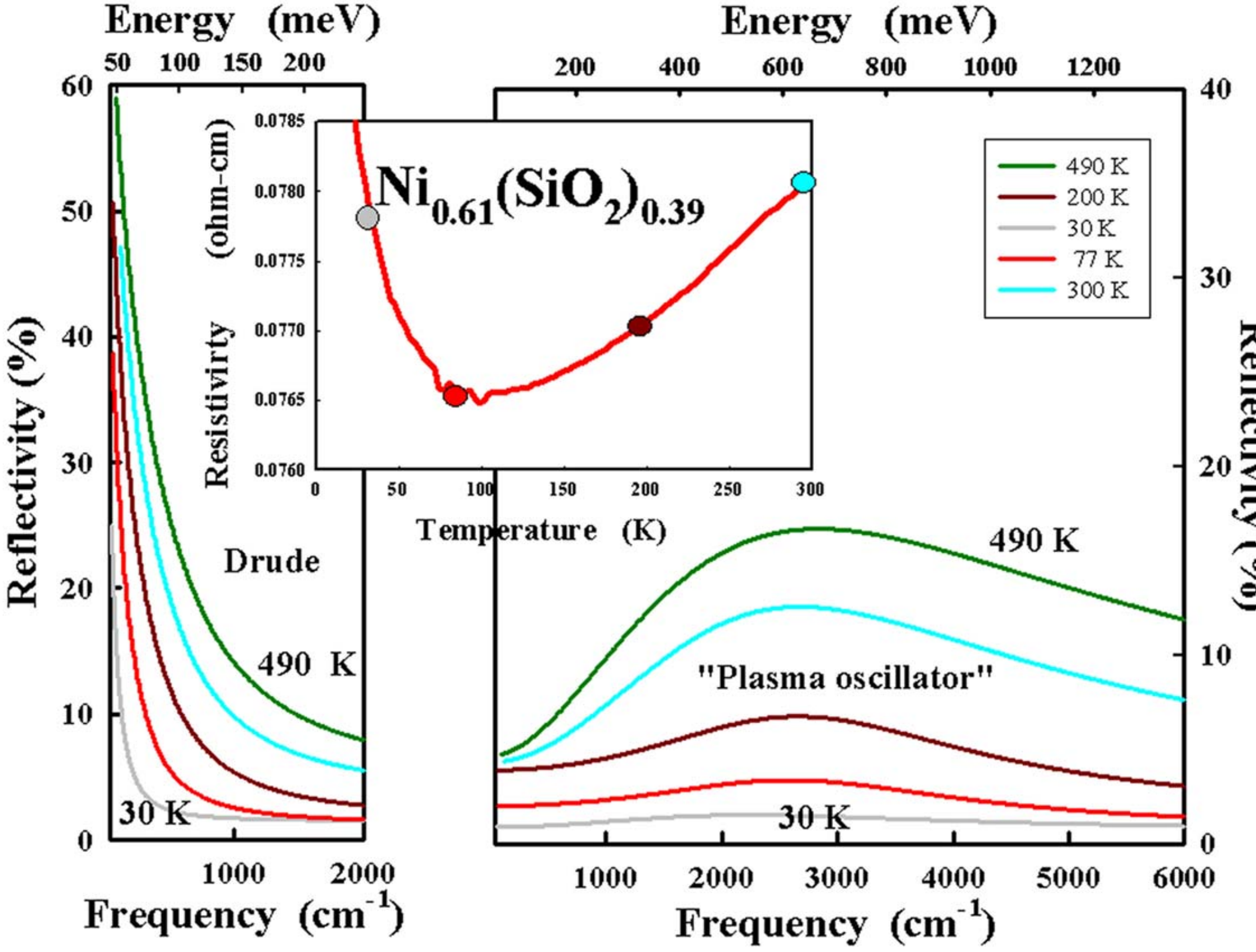

**Figure 7**
**Massa et al**

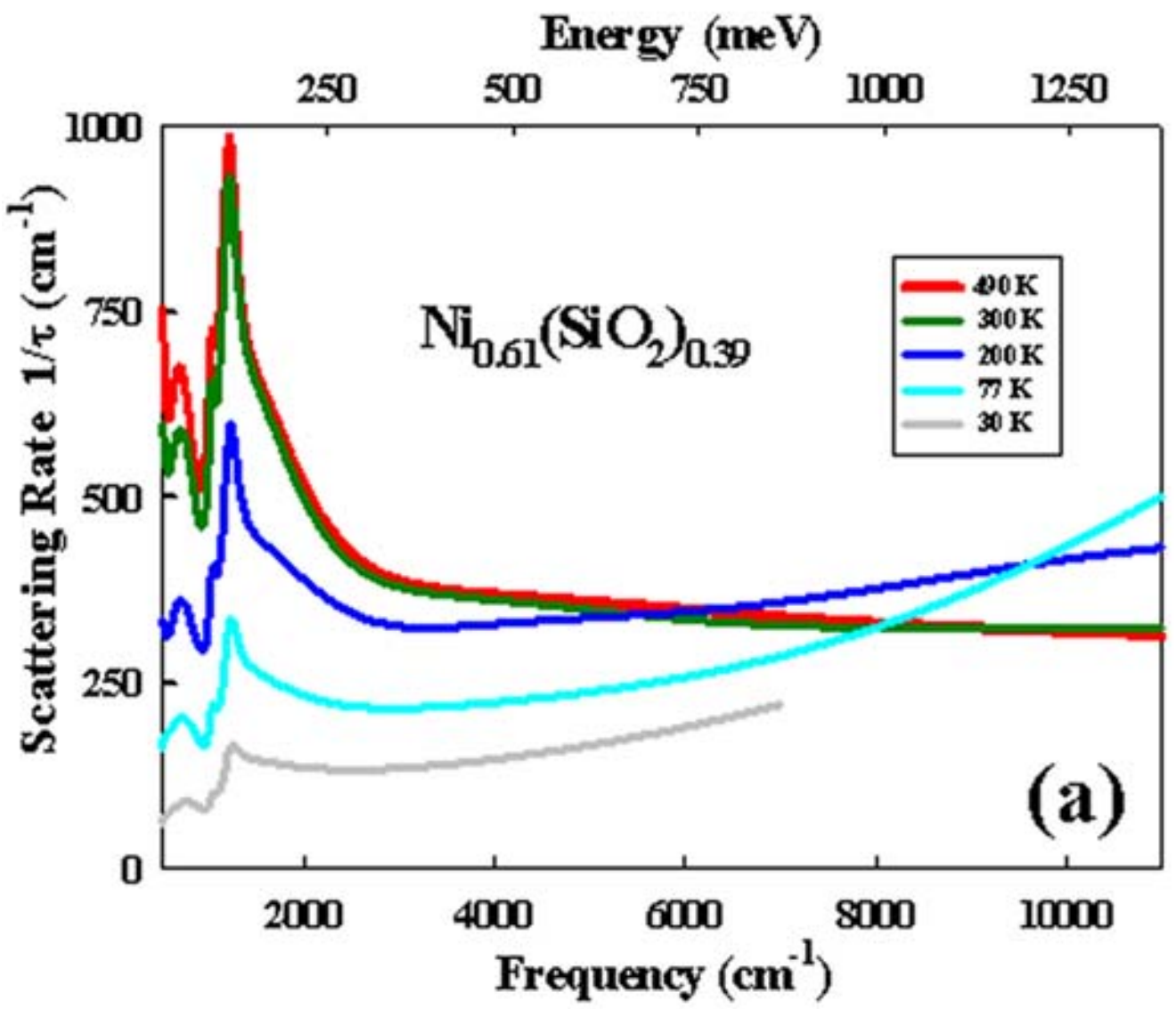

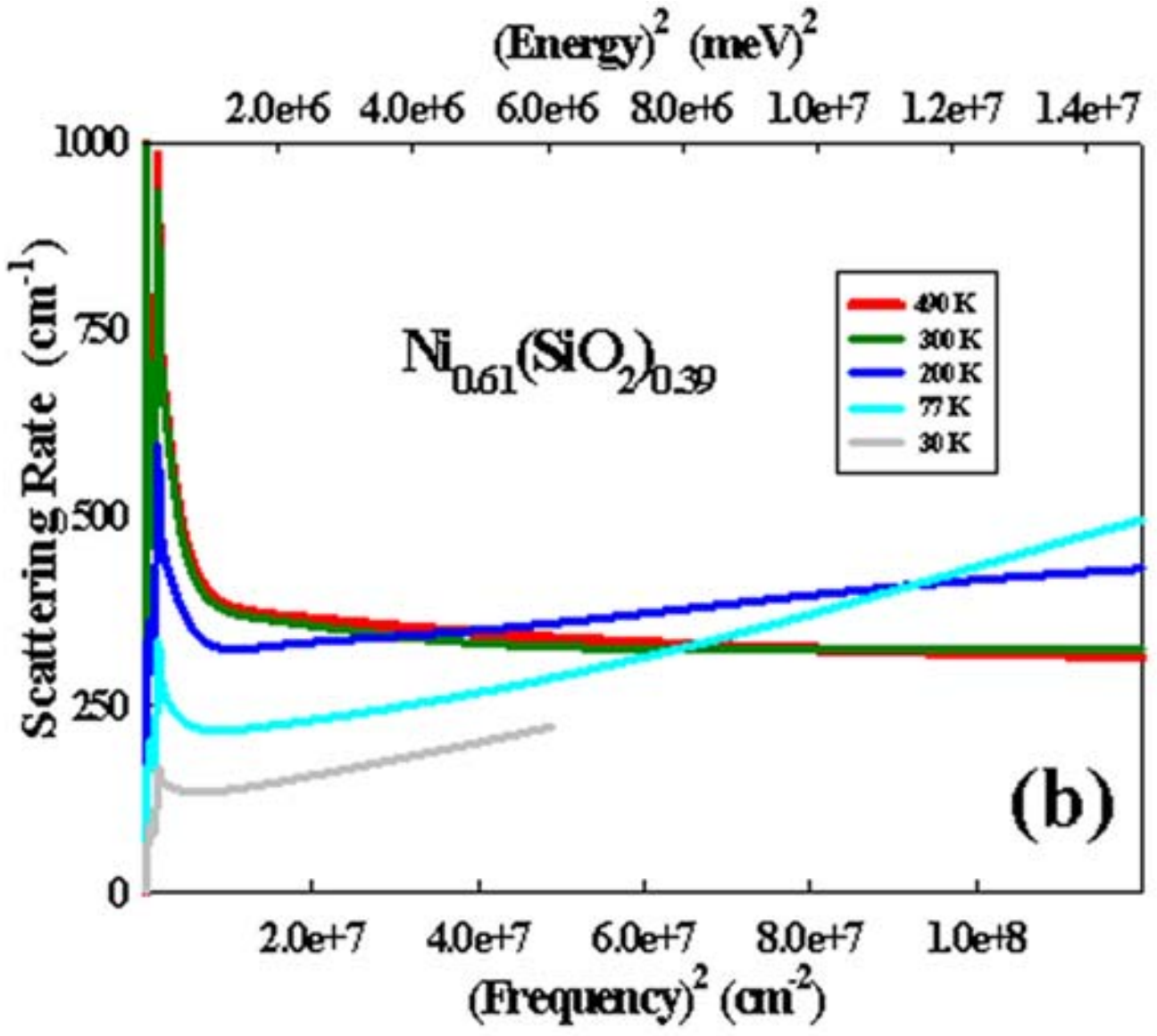

**Figure 8**
**Massa et al**

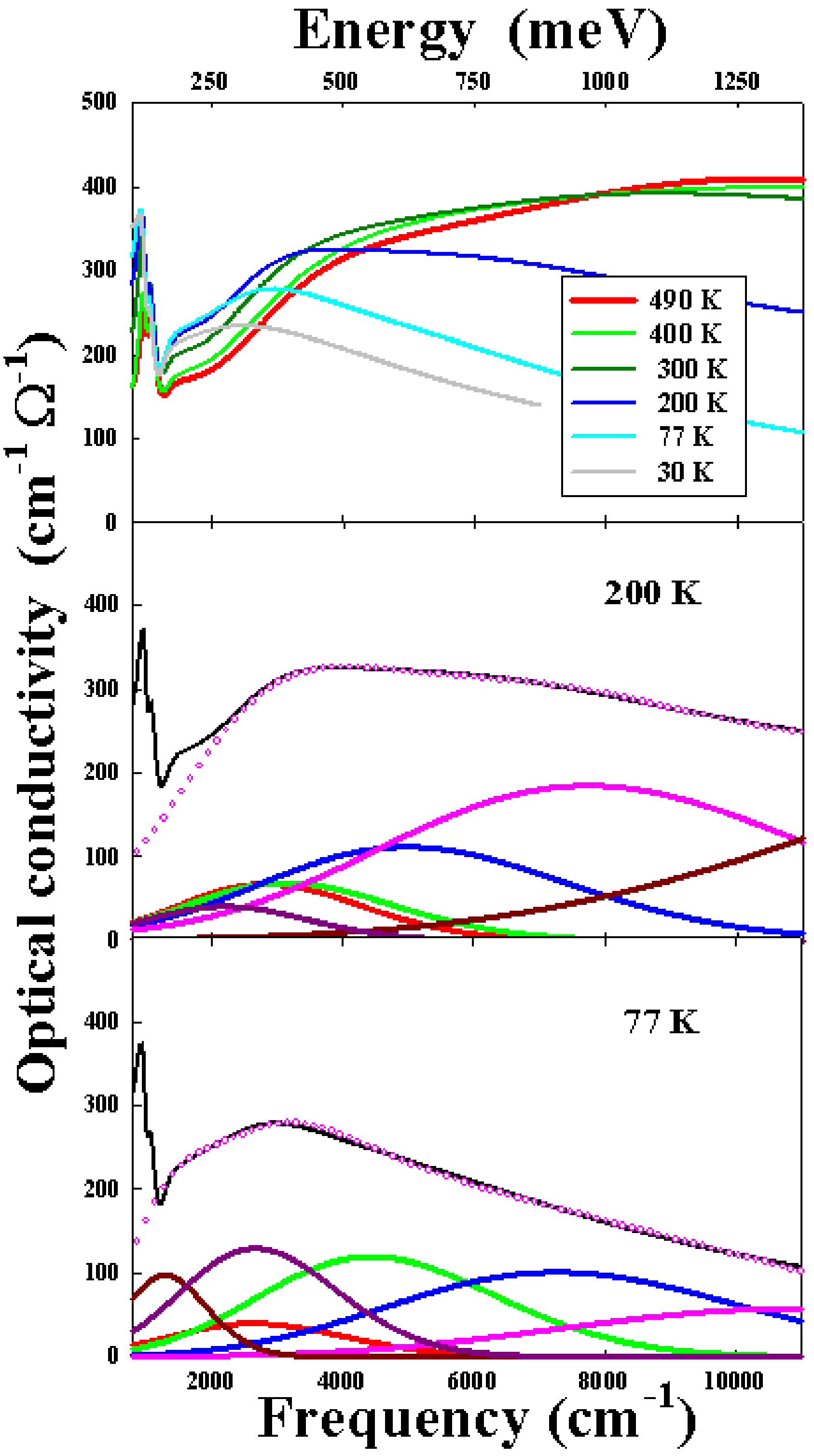

**Figure 9**
**Massa et al**

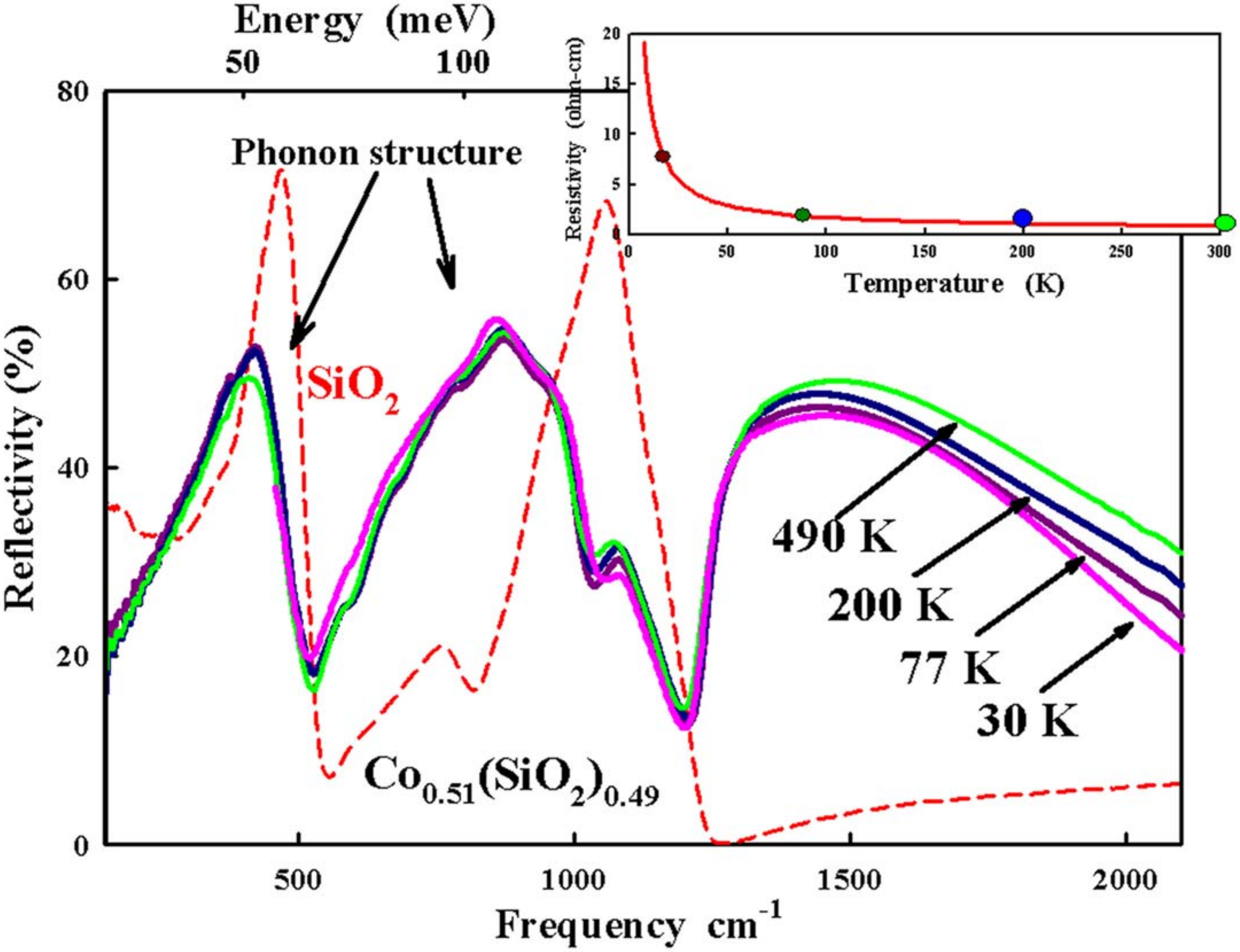

**Figure 10**
**Massa et al**

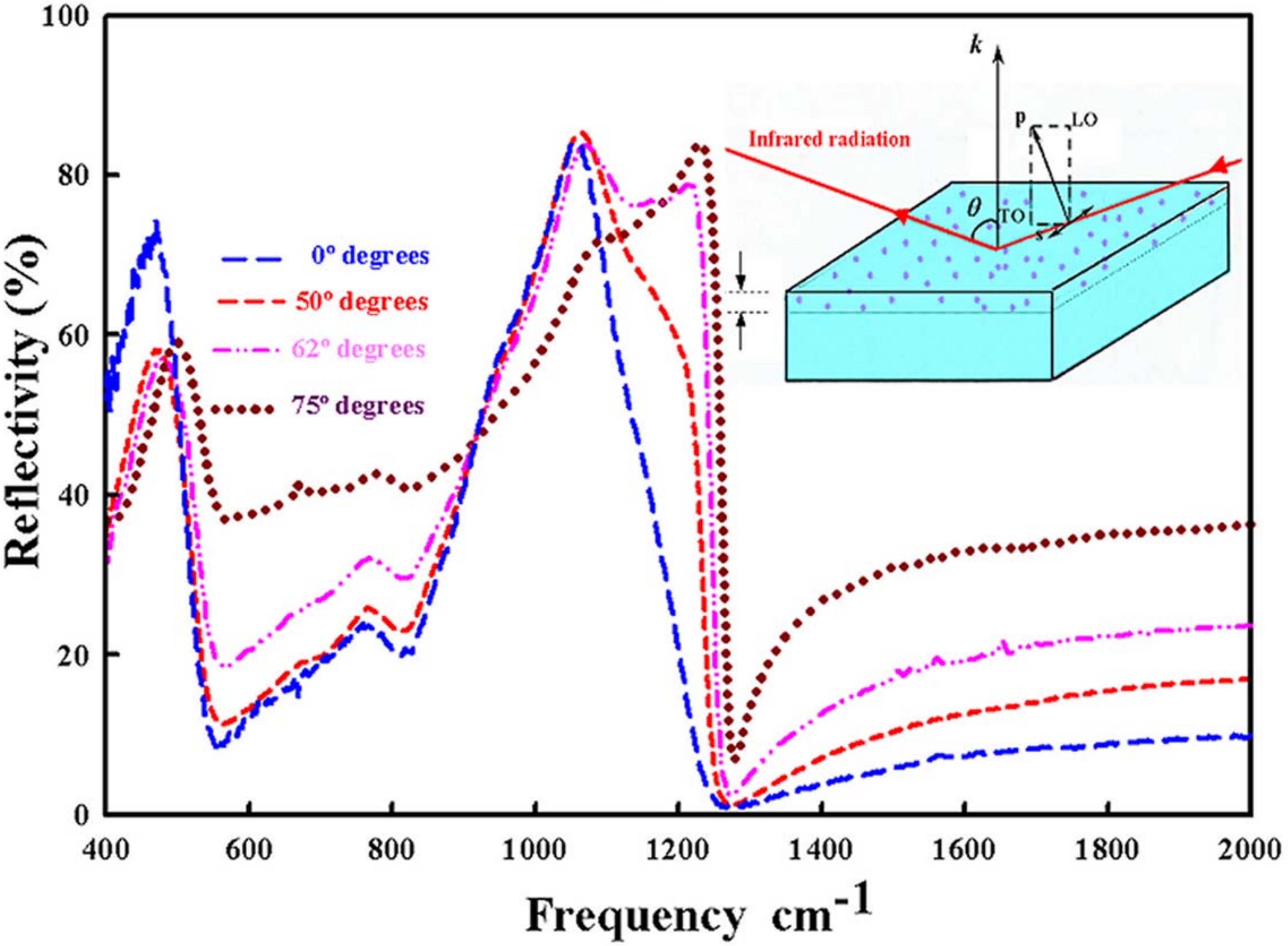

**Figure 11**
**Massa et al**

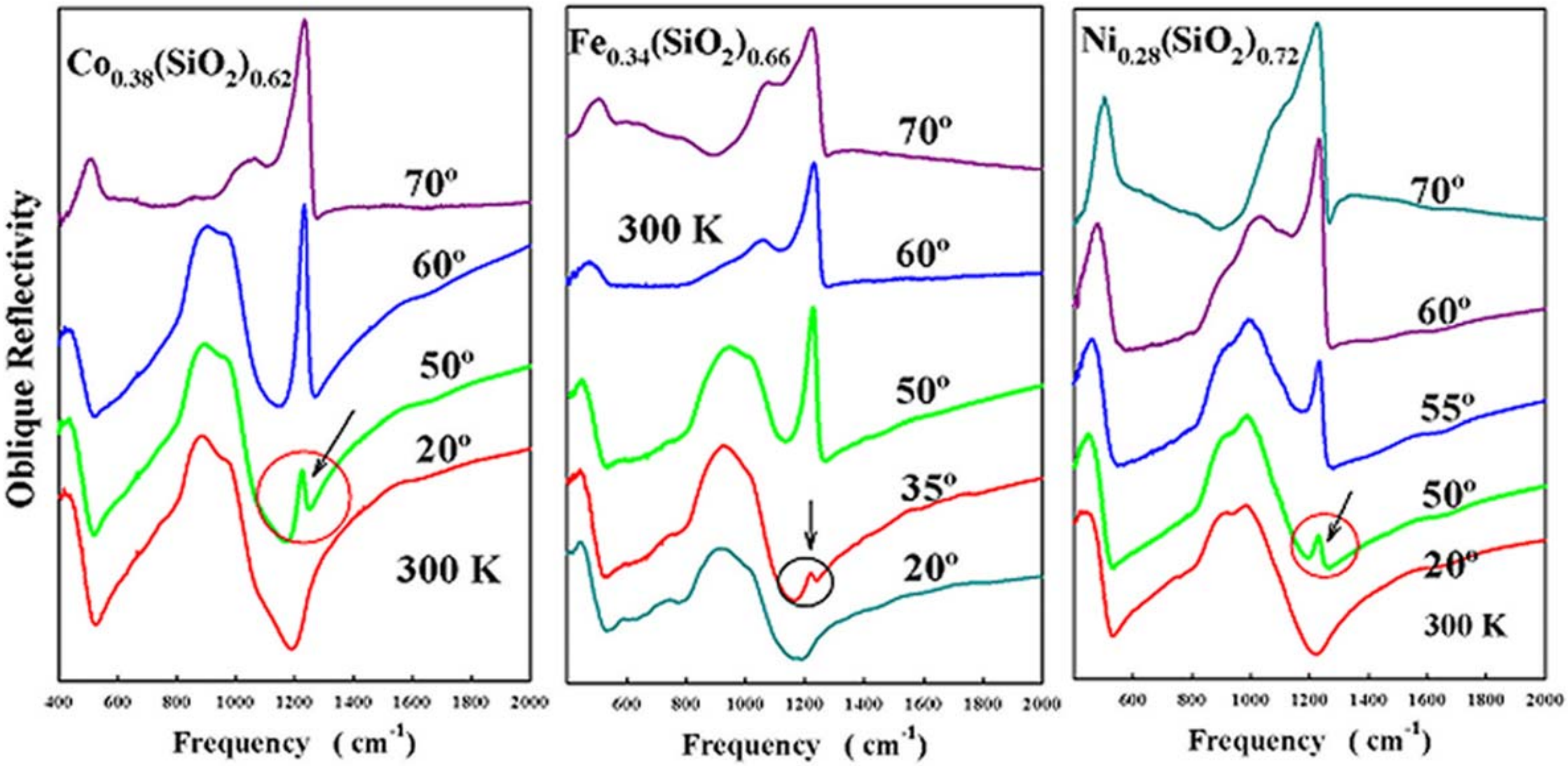

**Figure 12**
**Massa et al**

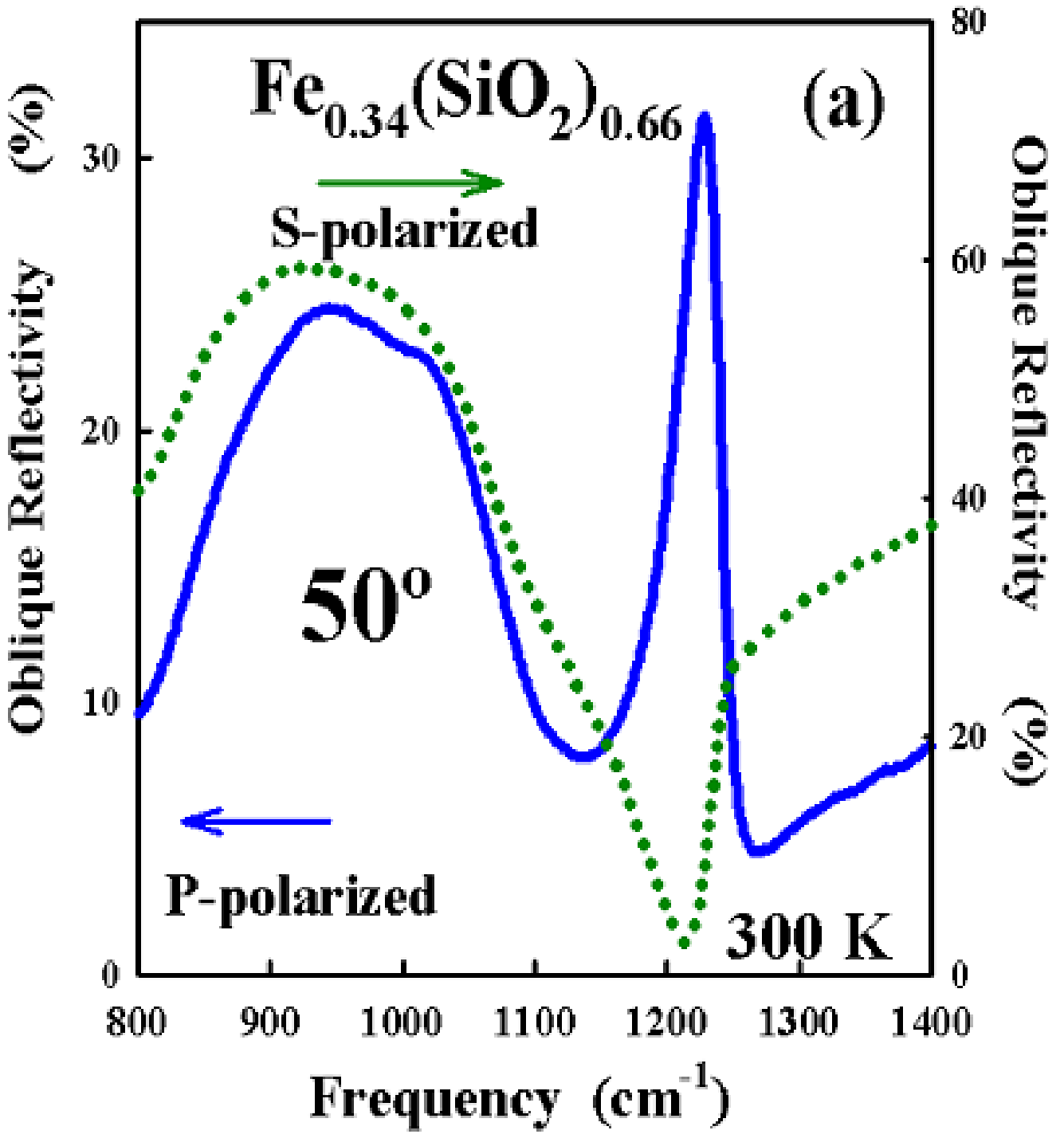

**Figure 13**
**Massa et al**

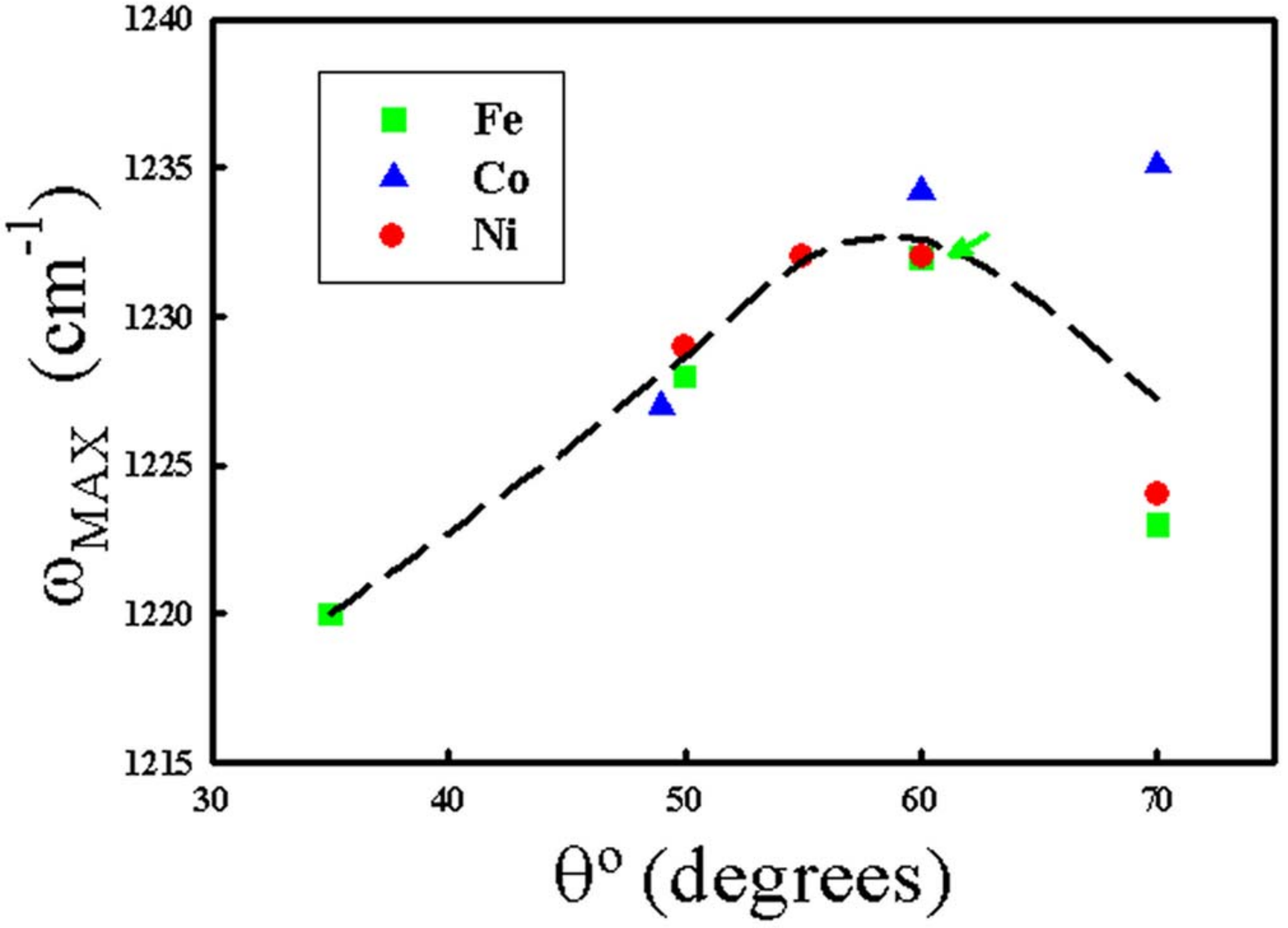

**Figure 14**
**Massa et al**

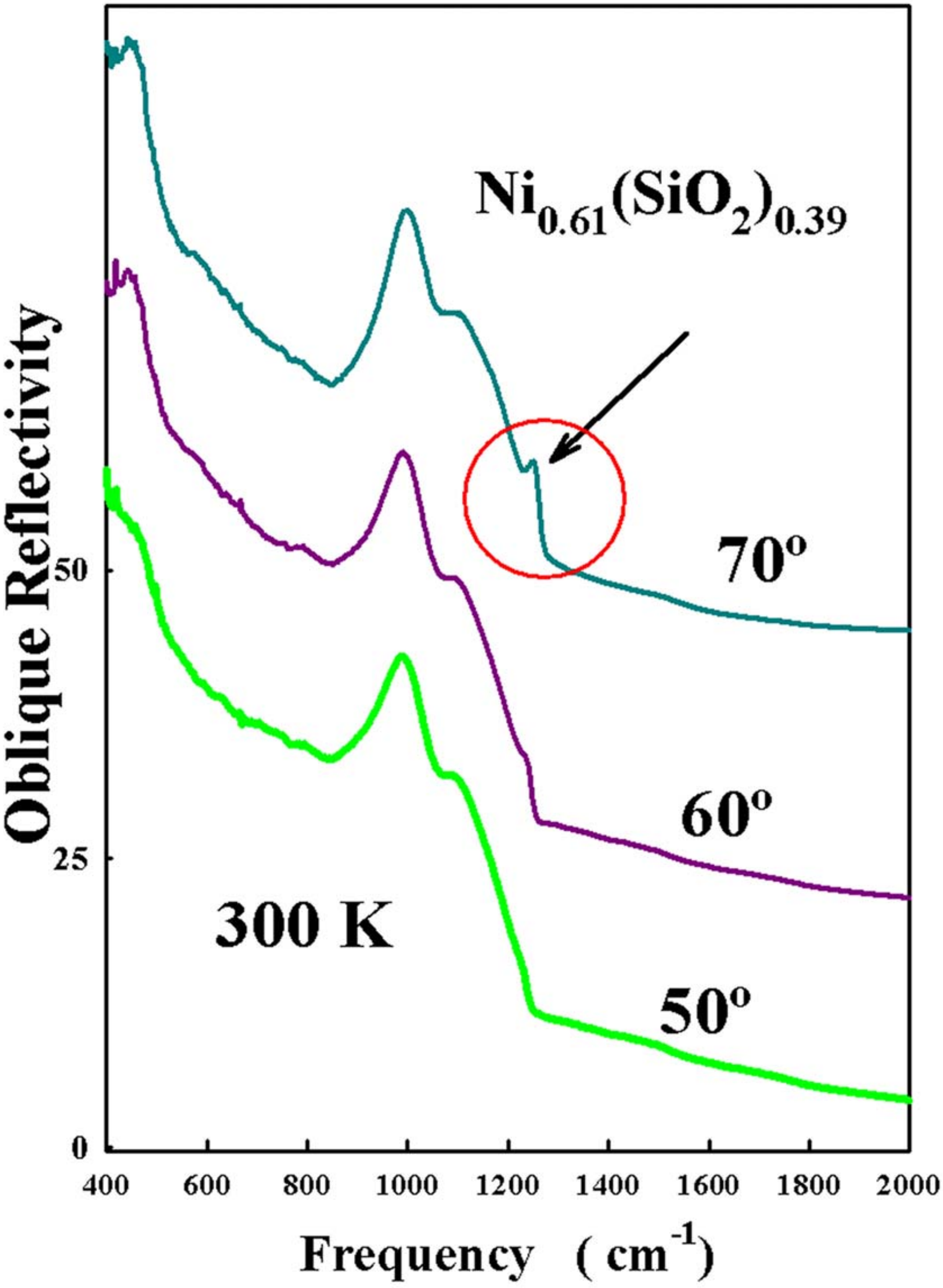

**Figure 15**
**Massa et al**